\documentclass[12pt,a4paper]{iopart}
\usepackage{iopams}
\usepackage{xspace}  
\usepackage{bm} 
\usepackage{epstopdf}
\usepackage[mathcal]{euscript}
\usepackage[latin1]{inputenc}
\usepackage{latexsym}
\usepackage{hyperref}
\usepackage{graphicx}   
\usepackage{verbatim}   
\usepackage{color}      
\usepackage{subfigure}  
\usepackage{hyperref}   
\usepackage{bm} 
\usepackage{amssymb}
\usepackage{bbm}
\usepackage{float}
\usepackage{slashed}
\usepackage{amsfonts}
\usepackage{mathrsfs} 
\usepackage{bbding}
\usepackage{pifont}
\def\r{\bm{r}}

\def\m{\bm{m}}
\def\n{\bm{n}}

\def\E{\bm{E}}
\def\B{\bm{B}}
\def\D{\bm{D}}

\def\J{\bm{J}}
\def\A{\bm{A}}

\def\p{\bm{p}}
\def\bfnabla{\bm{\nabla}}

\def\bfOmega{\bm{\Omega}}

\newcommand{\hh}{{_{\rm H}}}
\newcommand{\fermi}{{_{\rm F}}}
\newcommand{\lorentz}{{_{\rm L}}}
\newcommand{\landau}{\lorentz}
\newcommand{\boltzmann}{{_{\rm B}}}
\newcommand{\pairing}{{_{\rm P}}}
\newcommand{\critical}{{_{\rm C}}}
\newcommand{\superfluid}{{_{\rm S}}}
\begin{document}
\title[Electromagnetism as an emergent phenomenon: a step-by-step guide]{Electromagnetism as an emergent phenomenon: a step-by-step guide}

\author{Carlos Barcel\'o$^1$, Ra\'ul Carballo-Rubio$^1$, Luis J.~Garay$^{2, 3}$, and Gil Jannes$^{4}$}

\address{$^1$ Instituto de Astrof\'{i}sica de Andaluc\'{i}a (IAA-CSIC), Glorieta de la Astronom\'{i}a, 18008 Granada, Spain}
\address{$^2$ Departamento de F\'{i}sica Te\'orica II, Universidad Complutense de Madrid, 28040 Madrid, Spain}
\address{$^3$ Instituto de Estructura de la Materia (IEM-CSIC), Serrano 121, 28006 Madrid, Spain}
\address{$^4$ Modelling \& Numerical Simulation Group, Universidad Carlos III de Madrid, Avda. de la Universidad 30, 28911 Legan\'{e}s, Spain}

\eads{
\mailto{carlos@iaa.es}, 
\mailto{raulc@iaa.es}, 
\mailto{luisj.garay@ucm.es},
\mailto{gil.jannes@uc3m.es}
}

\begin{abstract}

We give a detailed description of electrodynamics as an emergent
theory from condensed-matter-like structures, not only {\it per se}
but also as a warm-up for the study of the much more complex case of 
gravity. We will concentrate on two scenarios that,
although qualitatively different, share some important features, with
the idea of extracting the basic generic ingredients that give rise to
emergent electrodynamics and, more generally, to gauge theories. We
start with Maxwell's mechanical model for electrodynamics, where
Maxwell's equations appear as dynamical consistency conditions. We
next take a superfluid $^3$He-like system as representative of a broad 
class of fermionic quantum systems whose low-energy physics reproduces 
classical electrodynamics (Dirac and Maxwell equations as dynamical low-energy
laws). An important lesson that can be derived from both analyses is
that the vector potential has a microscopic physical reality and that
it is only in the low-energy regime that this physical reality is
blurred in favour of gauge invariance, which in addition turns out to
be secondary to effective Lorentz invariance.

\end{abstract}


\tableofcontents

\section{Introduction}

This  work is about emergent electromagnetism, but its motivation comes from trying to construct an emergent theory of gravity. Let us start the story from the beginning.

Given the classical behaviour of general relativity, it is unavoidable to think that there must exist some deeper-layer theory regularizing its singularities. The search for such a theory, generically denoted quantum gravity given the two main ingredients it is supposed to incorporate, is one of the corner stones of modern theoretical physics. Merging together quantum mechanics and general relativity encounters a number of difficulties, most of them arguably emanating from their different empathies towards the presence of background structures: whereas quantum mechanics is easily implementable when a background structure exists, general relativity demands the absence of an a priory fixed background structure. Given this situation, one can opt for trying to adapt quantum mechanics so as to elevate background independence, {\it i.e.} a geometrical viewpoint, to a fundamental principle ({\it e.g.} think of the loop quantum gravity approach~\cite{Rovelli2004,Gambini2011,Thiemann2007}). 

However, the situation has also led some researchers to ask themselves whether Einstein's theory could be just an emergent classical theory~\cite{Jacobson1995,Padmanabhan2010b,Padmanabhan2009}. From this perspective one does not have to strictly quantize general relativity, but to search for an underlying structure,  containing in principle no geometric notions whatsoever, such that classical general relativity can emerge at a coarse-grained level. In this work we will use the word emergent in this sense. For example, we would consider the string theory approach to quantum gravity as emergent, but approaches such as causal dynamical triangulations, causal sets or loop quantum gravity as non-emergent. Taking aside the much-developed string theory approach, there exist some  much less explored emergent-gravity approaches based on condensed-matter-like systems~\cite{Volovik2008,Konopka2008}. Contrarily to string theory, these latter approaches keep no relativistic trace at the fundamental level: even special relativity is emergent. 

In the last 10 years it has become clear that the appearance of metric structures controlling the propagation of effective fields within condensed matter systems, mostly in low-energy regimes, is quite simple and ubiquitous~\cite{Barcelo2001}. However, in general, these metric structures do not follow Einstein's equations, and it turns out to be very difficult to force them to, even at a theoretical level~\cite{Barcelo2001,Volovik2003}. It is not even clear what the fundamental origin of this difficulty is. In the context of an emergent dynamics {\it \`a la} Sakharov, this difficulty has been traced back to the ubiquitous non-relativistic behaviour of the effective fields at the scale playing the role of Planck scale in these systems~\cite{Volovik2008,Barcelo2010}.   

Given the difficulties in constructing a complete emergent theory of gravity within this setting, we decided to explore in detail all the steps involved in the construction of the much simpler case of emergent electromagnetism, with the idea of refreshing our minds before coming back to the gravitational problem. Moreover, to our knowledge, there does not exist a work of reference in which this construction in condensed matter systems (in particular, in Helium--3) is performed in a step-by-step fashion, making transparent all the hypotheses and approximations involved. It is our intention that this work may serve as a study guide for specialists in other approaches to quantum gravity as well.

Although our current experimental knowledge of quantum electrodynamics has not asked for a revision, it is still interesting to analyse the structure of a possible deeper layer underneath electrodynamics. From an exercising perspective, as we have said it is always helpful to understand simpler systems before embarking in more complicated endeavours. From a more physical perspective, there are partial emergent models that suggest that gravity and electromagnetism might emerge in a unified manner from a single underlying system~\cite{Volovik2008} and, indeed, this is also the situation in string theory \cite{Polchinski1998,Zwiebach2004}. If the very arena in which physics takes place, spacetime, has a discrete underlying structure, it is sensible to think that electromagnetism would also have such structure. 

In this work we present two models of emergent electromagnetism. One is originally due to Maxwell himself~\cite{Maxwell1861a,Maxwell1861b}. We revise and slightly update Maxwell's hydrodynamical model in the light of the physics we know today. The other model, which constitutes the bulk of the paper, is more sophisticated and is based on ideas coming from what we know about the superfluid phases of Helium--3. This construction follows the lead of the works of Volovik (see~\cite{Volovik2003} and references therein), and among other things intends to make his ideas more accessible to non-specialists in condensed matter. Many steps in the construction have our own perspective though, so that any misjudgement or error can only be blamed on us.   

Remarkably, Maxwell arrived to his unification of light and electromagnetism through the development of a mechanical model that could underlie all of the electromagnetic phenomena~\cite{Maxwell1861a,Maxwell1861b}. He imagined the electromagnetic aether as consisting of an anisotropic and compressible fluid made of cells, capable of acquiring rotation, separated by a layer of small idle wheels or ball bearings capable of rotating and moving between the cells. The bodies would be immersed in this fluid as an iron ball is immersed in water; they would distort the fluid around them. He did not commit  with this specific model as truly representing physical reality, though, but defended it on the grounds of a ``proof of principle'' of the possibility of formulating electromagnetism as a mechanical model.

At present we know that the physics of the microscopic world is controlled by quantum mechanical notions. The second model presented here assumes as fundamental a quantum-mechanical substratum (so that we do not discuss at all the possibility that quantum mechanics itself could also be emergent). Then, it is developed so as to show how, at least classical, electrodynamics can emerge. Again, we consider this a proof of principle and not a commitment with the very form of the underlying physics. The analysis proceeds in a step-by-step basis. We try to discern which are the basic ingredients that would be common to any emergent theory of electromagnetism and which ones appear as particular of this specific construction.     

It has been argued that the most crucial step made by Maxwell was to abandon his mechanical model and just worry about the properties of the resulting coarse-grained effective field theory~\cite{Longair2003,Dyson1999,Simpson1997}. The field-theoretical point of view has since been a central theme in most developments in fundamental physics. Whereas nobody can deny the tremendous power and successes of this approach, it assumes many ingredients as a matter of principle, without a deeper explanation: {\it e.g.} Why is there a maximum velocity for the propagation of signals? Why is there gauge invariance? Why are elementary particles within a class indistinguishable? Why are there no magnetic monopoles? As we will see, an emergent approach is capable of providing explanations for many of these questions. On the other hand, an emergent perspective puts a stronger accent on the universal characteristics of possible microscopic theories than on the specifics of a particular implementation. We think that the emergent approach complements the field-theoretical approach, together providing a much richer source of understanding.

\section{An updated Maxwell fluid model}\label{sec:maxwell}

In Maxwell's time people did not have a clear idea of what electric currents really were, not to mention the then unknown atomic structure of matter. Given the present knowledge, we can propose an updated fluid model for electromagnetism following closely Maxwell's proposal \cite{Maxwell1861a,Maxwell1861b}. For other modern viewpoints on Maxwell's hydrodynamical model, see {\it e.g.} \cite{Moyer1978,Cat2001,Siegel1992,Simpson2010,Rousseaux2002,Rousseaux2005}.

Imagine a fluid made of two different elementary constituents: vortical cells and small ball bearings. A vortical cell is made of a topologically spherical and deformable membrane filled with a fluid. The details of this fluid are not very important in what follows so, to simplify matters, let us take it to be incompressible and highly viscous. The membrane provides a fixed constant tension in all its points. It supports tangential as well as normal tensions. In the case in which the membrane were put to rotation around an axis, the filling fluid would rapidly end up rotating with a uniform angular velocity around that axis. The total angular momentum of the vortical cell will be $I\Omega$, with $I$ its moment of inertia and $\Omega$ its angular velocity, or $Iv_{\rm e}/r_{\rm e}$, with $r_{\rm e}, v_{\rm e}$ its equatorial radius and velocity.

The fluid inside the cell has an isotropic (hydrostatic) pressure. When it is non-rotating this pressure is a constant $p_0$ throughout the cell. 
However, rotation provides centrifugal forces which change the pressure pattern.  
In the equator the pressure will have an excess $p_0+{1 \over2}\rho v^2$ with respect to the poles. Independently of the precise form of the cell,  the cell as a point will exert a pressure excess in the directions orthogonal to the axis, and this pressure excess will be proportional to the rotation velocity squared:
\begin{eqnarray}
p_\parallel = p_0 + C_\parallel \Omega^2~,~~~~
p_\perp = p_0 + C_\perp \Omega^2~,~~~~ C_\parallel < C_\perp ~.
\end{eqnarray}
We can also write this excess as
\begin{eqnarray}
\Delta p= p_\perp - p_\parallel= \mu_{\rm micro}^{-1} B_{\rm micro}^2~.
\end{eqnarray}
At this stage the dimensions of $\B_{\rm micro}$ and $\mu_{\rm micro}$ are not fixed but only  the dimensions of the above product. For later convenience let us choose $\B_{\rm micro}$ to denote minus the average density of angular momentum in the vortical cell, multiplied by a typical length scale in the system, $\B_{\rm micro}= -(I \bfOmega /V)R$. Then, the quantity $\mu_{\rm micro}$ is a constant with units $\{ML\}$ (mass times length). Although the dimensions have been fixed one can still multiply $\mu_{\rm micro}$ by a dimensionless number $N$ and $\B_{\rm micro}$ by $\sqrt{N}$ with no effect, or in other words, one can change the length scale $R$ that defines $\B_{\rm micro}$ if one redefines $\mu_{\rm micro}$ accordingly. One could also have defined $\B_{\rm micro}$ with a reversed sign with no effect (in fact, we have chosen the negative sign for later convenience). A specific definition of $\B_{\rm micro}$ will appear only when fixing an operational meaning for it. Let us advance here that when later introducing the unit of charge, it will be natural to define $\B_{\rm micro}$ with units $\{(J/L^3)(L/Q)\}$ and $\mu_{\rm micro}$ with units $\{ML/Q^2\}$ ($J$ is the angular momentum and $Q$ the charge). 

\begin{figure}
	\begin{center}
		\includegraphics[width=14cm]{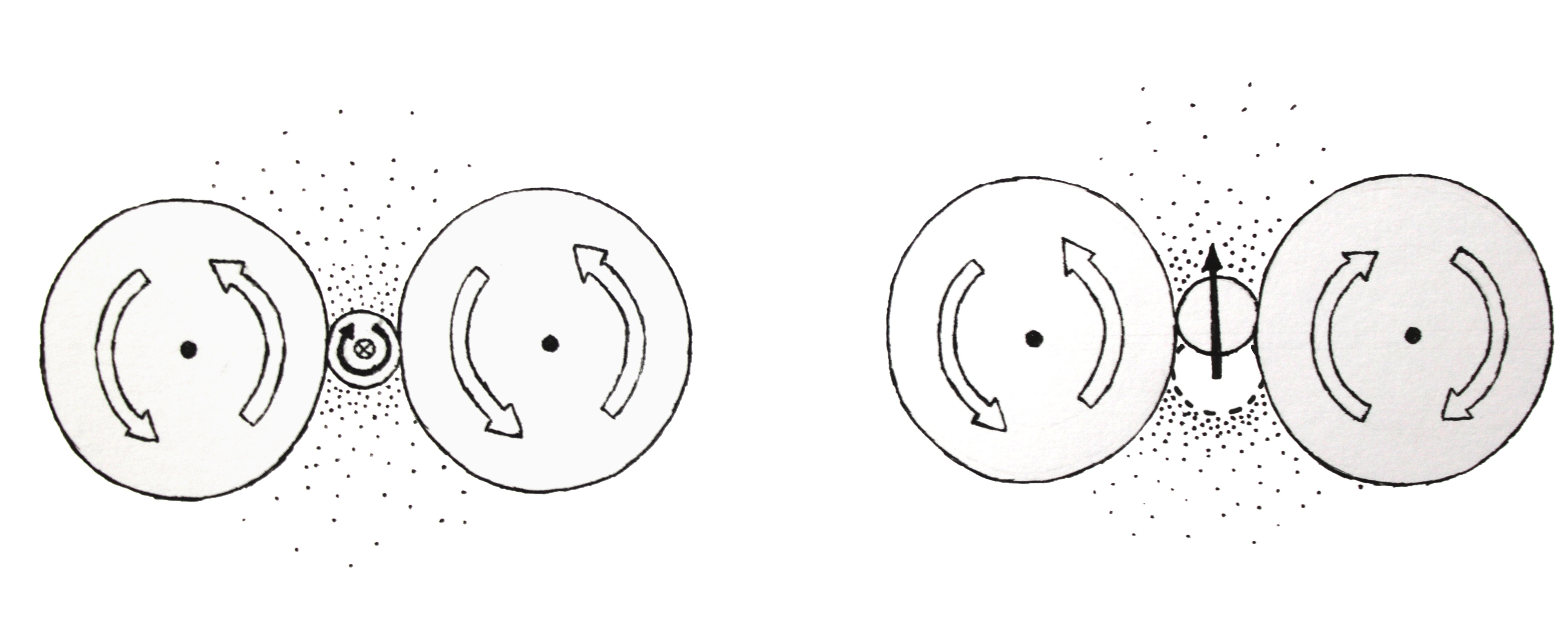} 
	\caption{Diagram showing the transfer of rotation between the different elements in the fluid.} \label{Fig:tension}
	\end{center}
\end{figure}

On the other hand, a ball bearing is a small spherical ball (much smaller than a vortical cell) that sticks to any membrane in such a way that whereas it can move over it, it cannot slide, that is, any movement has to be accompanied either by rotation or by a tangential stretching of the membrane itself. This fluid of small balls is also endowed with a hydrostatic (isotropic) pressure. This pressure produces a displacement of the microscopic distribution of ball bearings with respect to the vortical cells that in turn produces microscopic restoration forces. This is due to the fact that most ball bearings will be attached to at least two vortical cells so that the only way to move them is by creating a tangential distortion (and a subsequent tension) on the membranes (see Figure~\ref{Fig:tension}). Thus, the hydrostatic pressure of the ball bearings combined with their stickiness results in an equilibrium state endowed with tensions, which we will call, in a modern language, ``vacuum state''.

The complete description of a fluid made of a huge number of vortical cells with an even larger number of ball bearings stuck to their surfaces, all put together in a box, will be tremendously complicated and uncontrollable in practice. However, from a coarse-grained perspective, we could use just a few macroscopic variables to characterize the state of the fluid, as it is done in standard fluid mechanics. Consider one small part of the fluid but still containing a large number of constituents. At any coarse-grained point the vortical cells will contribute with an overall hydrostatic pressure $p_{\hh}$ plus some tension acting in a specific direction, the overall rotation axis. This will lead to an anisotropic pressure that can be written
\begin{eqnarray}
p_{ij} = \delta_{ij} p_{\hh} - \mu_0^{-1} B_i B_j~. 
\end{eqnarray}
Here, the vector $\B$ is the macroscopic version of $\B_{\rm micro}$ and therefore is proportional to the angular momentum density (total angular momentum in the coarse-grained point divided by its volume). The quantity $\mu_0$ is a constant with units $\{ML\}$. The same redefinition ambiguities associated with the microscopic quantities apply to their macroscopic versions. These $p_{\hh}$ and $\B$ are our first macroscopic variables.

Now, non-vacuum states can have tangential displacements of the ball bearings (with their associated restoring tensions) beyond their equilibrium positions. We can characterize these tensions by a microscopic displacement vector field $\D_{\rm micro}$ (displacement of each ball bearing with respect to its vacuum position). At the coarse-grained level we can construct a displacement-density vector field $\D$ and associated with it a force field, $\E=\epsilon_0^{-1}\D$ with $\epsilon_0$ for now a free constant with the appropriate dimensions. The real restoration force field will be proportional to the displacement and hence to this force field $\E$.

To recover (classical) electrodynamics from a fluid system like the one being described, we still need one more ingredient: Something has to play the role of charge. In the vacuum state, ball bearings are all strongly stuck to vortical cells so that they cannot move from one vortical cell to another. 
However, out of this vacuum state, there can be movable ball bearings able to performing macroscopic displacements jumping from cell to cell. To introduce movable ball bearings in the system one could even break some of the strong links of the ball bearings characterizing the vacuum. In this way ball bearings can be relocated in space. As we will see, the presence of regions with an overdensity or an underdensity of movable ball bearings with respect to the vacuum state can be associated with a positive and a negative charge density, respectively. These overdensities and underdensities will in turn be responsible for the redistribution of tensions in non-vacuum states that we described before.

Now we are in a position to explain how Maxwell's equations can emerge from this fluid system. 

\begin{itemize}
\item[i)] Any rotational of the force field $\E$ will exert a torque that will increase the angular momentum of the vortical cells (and thus decrease $\B$ since we have defined it to be minus an angular momentum density). Once a specific meaning for $\B$ is given (recall that one can redistribute a constant dimensionless factor $N$ between $\B$ and $\mu$, or in other words, one has an initial flexibility in defining the length scale $R$), one can always find a specific $\epsilon_0$ so as to write 
\begin{eqnarray}
{\partial \B \over \partial t}=- \bfnabla \times \E~. 
\end{eqnarray}
In other words, this equation can be interpreted as fixing the value of $\epsilon_0$ with to respect $\mu_0$, and so fixing the relation between the
force field $\E$ and the displacement field $\D$.

\item[ii)] The presence of a ball-bearing overdensity or underdensity produces a change in the displacement field. Assuming the displacements to be sufficiently small, unless the system is unnaturally fine-tuned, we will always be able to write
\begin{eqnarray}
\bfnabla \cdot \D \approx  \rho_Q~ ~~\to ~ ~~\epsilon_0 \bfnabla \cdot \E \approx \rho_Q~.
\end{eqnarray}
This charge density $\rho_Q$ will be tightly related to the density of ball bearings with respect to the vacuum state. However, there is no need that they perfectly coincide. The only quantity with a macroscopic operational meaning (at least at this linear level) will be the charge density. At this stage one could introduce some reference unit of charge, and accordingly change the units of all the quantities by referring them to the effect of this reference charge.

\item[iii)] When the ball bearings move they exert torques on the cells. This applies to both ball bearings strongly stuck to the cells (not movable to other cells), which produce a change in the displacement field, and to ball bearings movable between cells (they are associated with charge currents). Reciprocally, when the rotation field of the vortical cells $\B$ acquires some rotational, it causes the ball bearings in the region to move (within their respective possibilities). We can encode this behaviour in an equation of the form   
\begin{eqnarray}
\bfnabla \times (\mu_0^{-1} \B)= \J_T~;~~~~ \J_T=\J_Q + {\partial \D \over \partial t}~.
\end{eqnarray}
The first term of the current $\J_T$ is due to the movable ball bearings (a proper current of charge) while the second term is due to the displacement of the non-movable cells (hence its name displacement current). This equation fixes the value of $\mu_0$ or, equivalently, the precise definition of $\B$ (the equation determines the value of the length-scale constant $R$). It is interesting to note that there is a curious interplay between the displacement current term and the occurrence of relativistic dynamics, as this is the term which is absent in the magnetic limit of Galilean electrodynamics \cite{Rousseaux2013}.

\item[iv)] Let us assume that the rotation field is divergenceless, although a priori there is no reason why this should be the case (more on this later). Then, we will have 
\begin{eqnarray}
\bfnabla \cdot \B =0~.
\end{eqnarray}

\end{itemize}
That it is, we have recovered all of Maxwell's equations from a mechanical fluid system. Let us make some observations here:
\begin{itemize}
\item
As already remarked by Maxwell, the important point here is not the specific details of this specific fluid model but the fact of its very existence. 
The equations for the macroscopic fields will not depend strongly on these details. In the derivation it has been necessary to make the assumption of smallness of all the perturbations with respect to the vacuum state. Beyond this regime one would expect to observe non-linear effects. For example, one would expect non-linear pressures of the form   
\begin{eqnarray}
p_{ij} = \delta_{ij} p_0  - \mu^{-1}(B,E) B_i B_j~. 
\end{eqnarray}
In the linear limit one can approximate $\mu(B,E)$ by a constant $\mu_0$.
\item
From Maxwell's equations one immediately deduces that this system admits light-like perturbations. These perturbations will move with a speed $c=\sqrt{\epsilon_0\mu_0}$. The fluid system can be perfectly described using Newtonian physics in which there is no limitation to the velocity of the bodies. Nonetheless, light speed shows up directly from the elastic properties of the body. The crucial ingredient for generalised sound velocities to emerge is that variations in time of local properties depend on local gradients of these same properties.
\item
Given that $\bfnabla \cdot \B=0$ one can always write $\B= \bfnabla \times \A$ locally.
For instance we could associate $\A$ with the macroscopic version of the flow lines of the fluid within the vortical cells. On the other hand, in places in which magnetic fields are stationary, $\bfnabla \times \E=0$ so that one can write $\E=-\bfnabla \phi$ locally. Then, the field $\phi$ represents a hydrostatic electric tension. A positively charged body will tend to move to places with smaller electric tension. In more general situations, due to the structure of Maxwell's equations, we can always write $\E=-\bfnabla \phi + \partial_t \A$. Knowing the coarse-grained structure of the fluid flow lines and hydrostatic electric tension, one knows $\phi$ and $\A$.   

Now, one can realize that regarding the values of $\E,\B$, the combination $\{\phi,\A \}$ and $\{\phi + \partial _t \chi, \A + \bfnabla \chi \}$ are equivalent. Within an emergence framework an appropriate interpretation of the previous condition, the gauge invariance condition, is that although the flow structure of the vortical cells and the electric tension both have a specific reality, different macroscopic configurations related through a gauge transformation are operationally indistinguishable from the effective dynamical theory (see also~\cite{Rousseaux2002,Rousseaux2005}). Gauge invariance appears because aspects of the system are ``invisible'' to observers restricted to experience only the effective fields.    

\item
In Maxwell's version of the fluid model, charges were associated directly with individual ball bearings, and currents with the movement of ball bearings from cell to cell. It seems completely unrealistic to have an electric current of this sort without some resistance or friction. But this resistance was perfectly accommodated in Maxwell's model by considering the currents as existing only within materials (conductors). Maxwell ascribed the ubiquitous resistance in a conducting wire to the collisions of the movable ball bearings when jumping between cells. However, this model will have a hard time to deal with a charged elementary particle in an otherwise empty space. There is no experimental evidence that the vacuum causes friction on a charge moving with uniform velocity. In fact, the presence of an effect of this sort would immediately uncover the existence of a privileged reference frame, against all we know about the relativity principle. 

However, we wanted our updated fluid framework to encompass also the movement of a free electron in an otherwise empty space. For that we have proposed to associate an electron to an overdense region of ball bearings (this identification is of course not complete as we have not attempted to specify the internal forces responsible for its structural stability). When thinking of an electron as a very localised overdense region, it appears difficult to find it in a uniform-velocity trajectory without dissipation of some sort. The movement of the ball bearings would be very noisy, with multiple collisions involved. However, if one imagines a pure plane-wave distribution of the overdense regions, it appears perfectly plausible for the propagation of the wave to occur without any appreciable friction: the propagation of the wave will not involve the presence of a macroscopic current of ball bearings. It is interesting to point out an analogy between this behaviour and that of free quantum particles. A position eigenstate of a wave function can propagate with a certain velocity, but as it travels it diffuses in space. This diffusion might be seen as an analogue in the quantum formulation of the noisy propagation expected in our fluid model. On the other hand, a momentum eigenstate is also an eigenstate of the Hamiltonian and it is not distorted by the propagation. 

It is also interesting to realize that the system allows in principle pair-creation of particles from the vacuum. If one pulls out ball bearings in one place and moves them to another region, one would have created equivalent overdense and underdense regions. The appearance of these ``quantum-like'' behaviours in our model might be interpreted as suggesting that the quantum mechanism itself could be an emergent phenomenon. Here we just mention this possibility without pursuing it any further.        
       
\item
The absence, as far as we know, of magnetic monopoles in nature has always being striking. 
As mentioned above, the fluid model a priori allows magnetic monopole configurations. However, when looking at the model carefully one realizes that this kind of configurations does not seem to be favoured by the system. Microscopically speaking, a magnetic monopole involves rotating cells with their angular momenta distributed radially. Any ball bearing located at the confluence of these cells will produce friction since the 
cells cause dragging forces incompatible with the no-sliding condition (see diagram in Figure~\ref{Fig:magnetic-monopole}). It seems reasonable that the system would tend to avoid these configurations (the same argument applies at the macroscopic level). The divergence-free condition $\bfnabla \cdot \B=0$ simply encodes the absence of these 
configurations.

\begin{figure}
	\begin{center}
		\includegraphics[width=10cm]{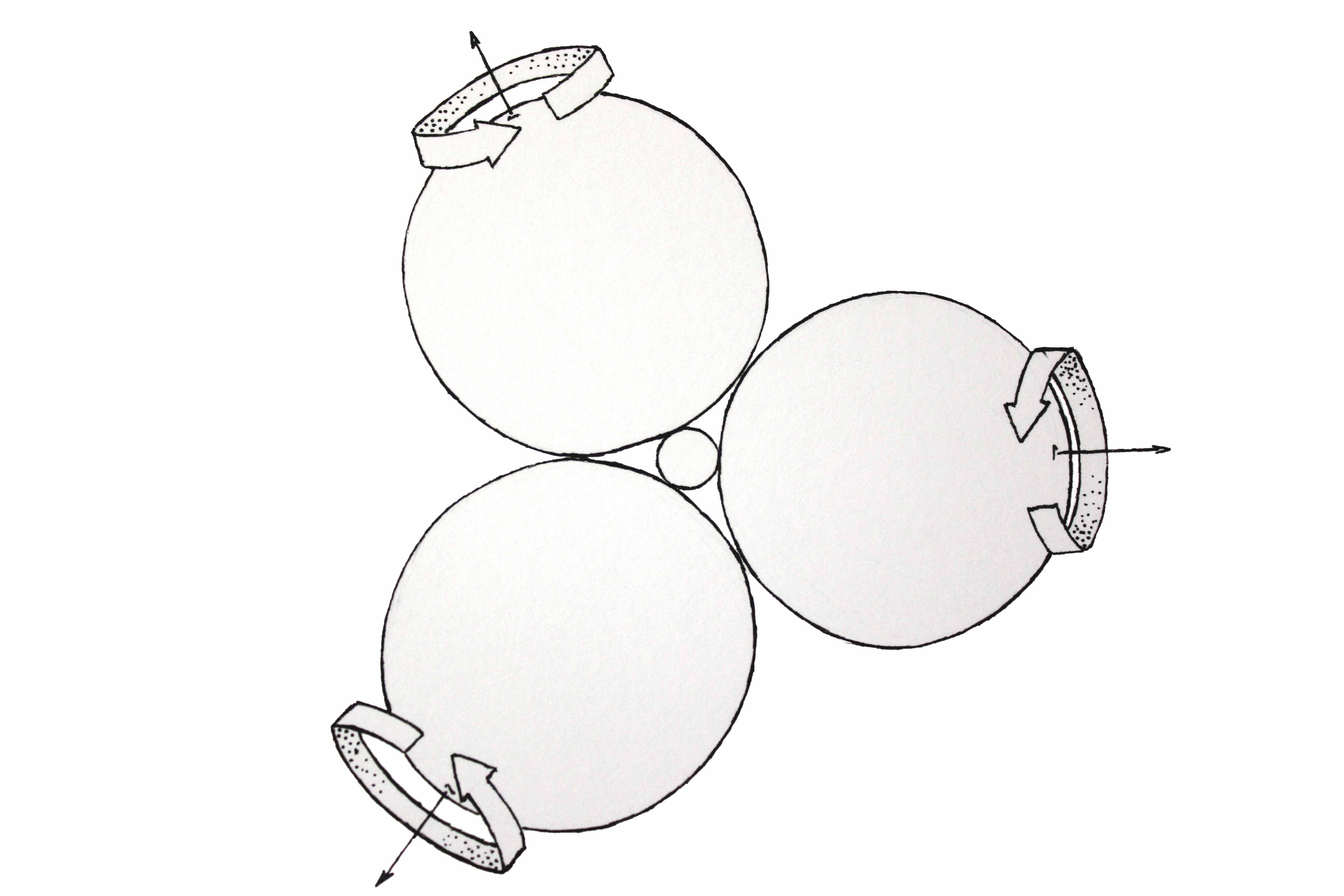}	
	\end{center}
	\caption{Diagram explaining the absence of magnetic monopoles. The system avoids these configuration because they would create friction for the central ball bearing at confluence of the rotating cells.} \label{Fig:magnetic-monopole}
\end{figure}

\item
It is interesting to estimate how small the constituents of this fluid have to be in order to pass unnoticed to current experiments. The smallest length scale ever tested is of the order of $10^{-19}$ m. So in principle a fluid structure several orders of magnitude beyond $10^{-19}$ m would remain undetected. Notice that the Planck length is $10^{-35}$ m, still 15 orders of magnitude ahead (equivalent to compare a human with interstellar distances).  

\item
It is also interesting to point out the different nature of light excitation and charged matter, even when considering it as wave-like. One can perfectly imagine a charge density with no overall velocity, representing a charged particle or distribution of particles at rest. Light is however similar to a phonon excitation of a lattice, it always travels with its fixed velocity (of course at very high energies one would expect to develop some dispersive effects). 
  
\end{itemize}

\section{Model based on Helium--3}

In this section we shall present a model of emergent electrodynamics based on the well-established theoretical understanding of the physics of Helium--3. Our presentation of the model will follow a top-down scheme. Nonetheless, we would like to stress the fact that these theoretical ideas were developed in close feedback with experiments and are proved to a great extent by them.  

Most of the introductory material covered here is well understood nowadays but 
as far as we know, it has not been presented in a logical step-by-step order so as to lead to a final emergent electrodynamics. In the following, only specific references are quoted. 

For a general discussion on superfluid Helium--3, one can draw on the review \cite{Leggett1975}, or the books \cite{Vollhardt1990,Leggett2006} and references therein. Concerning the low-energy properties of this system and analogies with other branches of physics, including relativistic field theories, the seminal reference is \cite{Volovik2003}.

\subsection{Microscopic $^3$He-like systems \label{sec:micro}}

Let us consider a quantum liquid composed of a large collectivity of spin-$1/2$ atoms. Here we use the word ``atom" to mean that these spins need not be elementary objects (they  need not be precisely $^3$He atoms either). We require the interactions between these atoms to be short-range but otherwise they can be very complicated, including higher-than-two-body effects. We also require that the two-body interactions be characterised by a potential of Lennard-Jones type (which is rotationally invariant) plus possibly some interaction term involving the spins. Interactions in $^3$He indeed possess these characteristics.  

To solve a system of this sort in full detail is beyond human capacities. We need simpler theories which serve as approximate models of the exact microscopic theory. A first step in this direction is provided by Landau's Fermi-liquid theory. This theory starts from the exact description of a free Fermi gas. In the free theory there appears the notion of Fermi surface and a notion of quasiatom and quasihole excitations. Landau's hypothesis is that generically (at least under certain conditions of temperature and pressure) the $N$-particle ground state and the spectra of quasiatom and quasihole excitations ({\it i.e.} the spectra in the surroundings of the Fermi surface) of the above strongly interacting theories are in adiabatic one-to-one correspondence with that of the free theory~\cite{Leggett2006}. Under this hypothesis we can use the same labels for these states. There exists some microscopic justification of Landau's hypothesis~\cite{Leggett1975}.

Now, regarding all the physics associated with low energies (vacuum state and excitations close to it), one can substitute the precise strongly interacting theory by an equivalent weakly interacting theory of quasiatoms. For instance, in second quantization language and in a momentum representation, one can write Landau's Grand Canonical Hamiltonian as 
\begin{eqnarray}
\hat{H}_\landau-\mu\hat{N} =&&\hspace{-2mm}
\sum_{\bm{p} \alpha} 
\left({p^2 \over 2m^*} -\mu \right)
\hat{a}_{\bm{p}\alpha}^\dagger \hat{a}_{\bm{p}\alpha}
\nonumber\\
&&\hspace{-2mm}
+\frac{1}{2}\sum_{\bm{p}\bm{p}'\alpha\beta}f(\p,\bm{p}',\alpha,\beta)\hat{a}_{\bm{p} \alpha}^\dagger \hat{a}_{\bm{p} \alpha}
\hat{a}_{\bm{p}' \beta}^\dagger \hat{a}_{\bm{p}' \beta}.
\label{eq:landau}
\end{eqnarray}
Here $\hat{a}_{\bm{p}' \beta}^\dagger,\ \hat{a}_{\bm{p}' \beta}$ are respectively creation and annihilation operators of quasiatoms with $\alpha,\beta$ representing the spin degree of freedom. Following standard conventions, in the rest of the text we will write $O$ instead of $\hat{O}$ for any operator, leaving this notation for unit vectors. The expression (\ref{eq:landau}) can be used in situations in which the number of quasiatoms (equal to that of atoms) is kept fixed. 
The chemical potential is $\mu=p_\fermi^2/(2m^*)$ with $p_\fermi$ the Fermi momentum and $m^*$ the effective mass of the quasiatoms (this mass does not need to coincide with the mass of the initial atoms; in $^3$He it is a few times smaller).
The function $f(\p,\bm{p}',\alpha,\beta)$ must be symmetric under the exchange $\p,\alpha\leftrightarrow\bm{p}',\beta$ and can be used to fit the specific interaction. Both $m^*$ and $f$ are in principle phenomenological quantities that depend on details of the microscopic interaction. This model Hamiltonian has proved to be very successful for example for the description of the normal phase of $^3$He, in the temperature range between 1K and $0.03$K. 

Now, there exist many systems that are different in the details of their interactions but are however undistinguishable from a low-energy point of view. They form part of the same low-energy universality class. These universality classes are characterised by topological properties of the vacuum state. Therefore, when working out a theory of emergent electromagnetism, one is really obtaining a family of theories with the same low-energy behaviour. 
The same operators $a_{\bm{p}\alpha}$ can represent different physical quasiatoms in different strongly-interacting spin-fluid systems. These operators can also represent the proper atoms of a weakly-interacting spin-gas system. 
In the following, we will analyse the properties of a specific weakly interacting
theory, independently of any specific physical realization one could have in mind.
Thus, we will speak only of atoms, having always in mind that they could be equivalently quasiatoms.

\subsection{A weakly interacting gas \label{sec:weakly}}

Let us focus on a weakly interacting theory of spin-$1/2$ atoms.
One can go one step further than Landau's Fermi-liquid theory and analyse a more general interaction.

One can introduce the atom field $\psi$. Then, in second quantization, the Grand Canonical Hamiltonian for the system of spin-$1/2$ atoms with two-body interactions reads
\begin{eqnarray}
H-\mu N:&=\int\mbox{d}^3x\,{\psi}^\dagger(\bm{x})\left(-\frac{\hbar^2}{2m^*}\nabla^2-\mu\right){\psi}(\bm{x})
\nonumber\\
&+\frac{1}{2}\int\mbox{d}^3x\mbox{d}^3x'V(\bm{x}-\bm{x}'){\psi}^\dagger(\bm{x}){\psi}^\dagger(\bm{x}'){\psi}(\bm{x}'){\psi}(\bm{x})\label{eq:mbham}.
\end{eqnarray}
We have assumed for the time being that the interaction potential does not depend on the spin. In the momentum representation, ${\psi}_\alpha = \sum_{\p} {a}_{\bm{p}\alpha}e^{i\bm{p} \cdot \bm{x}}$, we have 
\begin{eqnarray}\fl
H-\mu N:=
\sum_{\bm{p} \alpha} 
\left({p^2 \over 2m^*} -\mu \right)
{a}_{\bm{p}\alpha}^\dagger {a}_{\bm{p}\alpha}
\nonumber\\
+\frac{1}{2} \sum_{\p_1 + \p_2 = \p_3 +\p_4, \alpha \beta} \tilde{V}\left(\frac{\p_1-\p_2 +\p_3 -\p_4}{2}\right) {a}_{\p_4 \beta}^\dagger {a}_{\p_3 \alpha}^\dagger
{a}_{\p_2 \alpha} {a}_{\p_1 \beta}
\label{eq:mbhammom},
\end{eqnarray}
with
\begin{eqnarray}
\tilde{V}(\p):=  \frac{p_\fermi^3}{\hbar^3} \int\mbox{d}^3r e^{i \p \cdot \r/\hbar} V(\r)\quad\mbox{and}\quad V(-\r)=V(\r)~.
\end{eqnarray}
We have taken this definition so that $\tilde{V}$ has dimensions of energy. Our notation in what follows assumes a finite volume and so a discrete sum in momentum space; the infinite volume limit can be obtained by adding the appropriate dimensionful constants. Notice that the potential term in (\ref{eq:mbhammom}) is invariant under a Galilean boost transformation of the reference frame. We should keep in mind this property, which can apparently be lost under certain approximations that will be made in the following.
This sum contains different interaction channels: the Hartree channel [which contains the previous Landau terms, Eq. (\ref{eq:landau})], the Fock channel, and the pairing channel~\cite{Leggett2006}. Of special relevance in what follows is the pairing channel that appears for interactions satisfying $\p_1=-\p_2=:\p$ and $\p_3=-\p_4=:\bm{p}'$. The pairing terms  control the form of the vacuum state of the theory (see Leggett's discussion in~\cite{Leggett2006}). The pairing Hamiltonian reads
\begin{eqnarray}
H_\pairing -\mu N:=&
\sum_{\bm{p} \alpha} 
\left({p^2 \over 2m^*} -\mu \right)
{a}_{\bm{p}\alpha}^\dagger {a}_{\bm{p}\alpha}
\nonumber\\
&+\frac{1}{2}\sum_{\bm{p} \bm{p}' \alpha \beta} \tilde{V}[(\bm{p}' +\p)] {a}_{-\bm{p}' \beta}^\dagger {a}_{\bm{p}' \alpha}^\dagger
{a}_{-\bm{p} \alpha} {a}_{\bm{p} \beta}. 
\label{eq:hampairing1}
\end{eqnarray}
If the potential does not depend on the orientation, it can only depend on $|\bm{p}'+\p|=p^2+p'^2+2pp'\hat{\bm{p}} \cdot \hat{\bm{p}}'$. Then, we can always write it as an expansion of the form \cite{Leggett1975}
\begin{eqnarray}
\tilde{V}(|\p+\bm{p}'|)= \sum_l\tilde{V}_l (p,p') P_l(\hat{\bm{p}} \cdot \hat{\bm{p}}')~,
\end{eqnarray}
where $P_l$ represents Legendre polynomials (the converse is not true: not all expansions can be put in exact correspondence with $V(r)$ potentials). As we are always interested in the surroundings of the Fermi surface (where the low-energy excitations reside), we can take the potential to depend mainly on the angle $\hat{\bm{p}} \cdot \hat{\bm{p}}'$ and not in the norms which will be $p,p'\simeq p_\fermi$.

Take now a microscopic interaction such that $|\tilde{V}_1| \gg |\tilde{V}_{l \neq 1}|$. Then $g:= -\tilde{V}_1 (p_\fermi,p_\fermi')$ will be a positive constant because of the binding character of the potential. The potential will be written as 
\begin{eqnarray}
\tilde{V} \simeq - g \hat{\bm{p}} \cdot \hat{\bm{p}}' \simeq - {g \over p_\fermi^2} \bm{p} \cdot \bm{p}'~.\label{eq:foucomp}
\end{eqnarray}
For instance, the simplest interaction of this kind is the one provided by 
\begin{eqnarray}
V(\bm{x}-\bm{x}')=-{g \over {8p_\fermi^2}} (\bm{\nabla}-\bm{\nabla}')^2 \delta(\bm{x}-\bm{x}')+g\delta(\bm{x}-\bm{x}'). 
\label{dpotential}
\end{eqnarray}

\begin{figure}[h]
	\begin{center}
		\includegraphics[width=10cm]{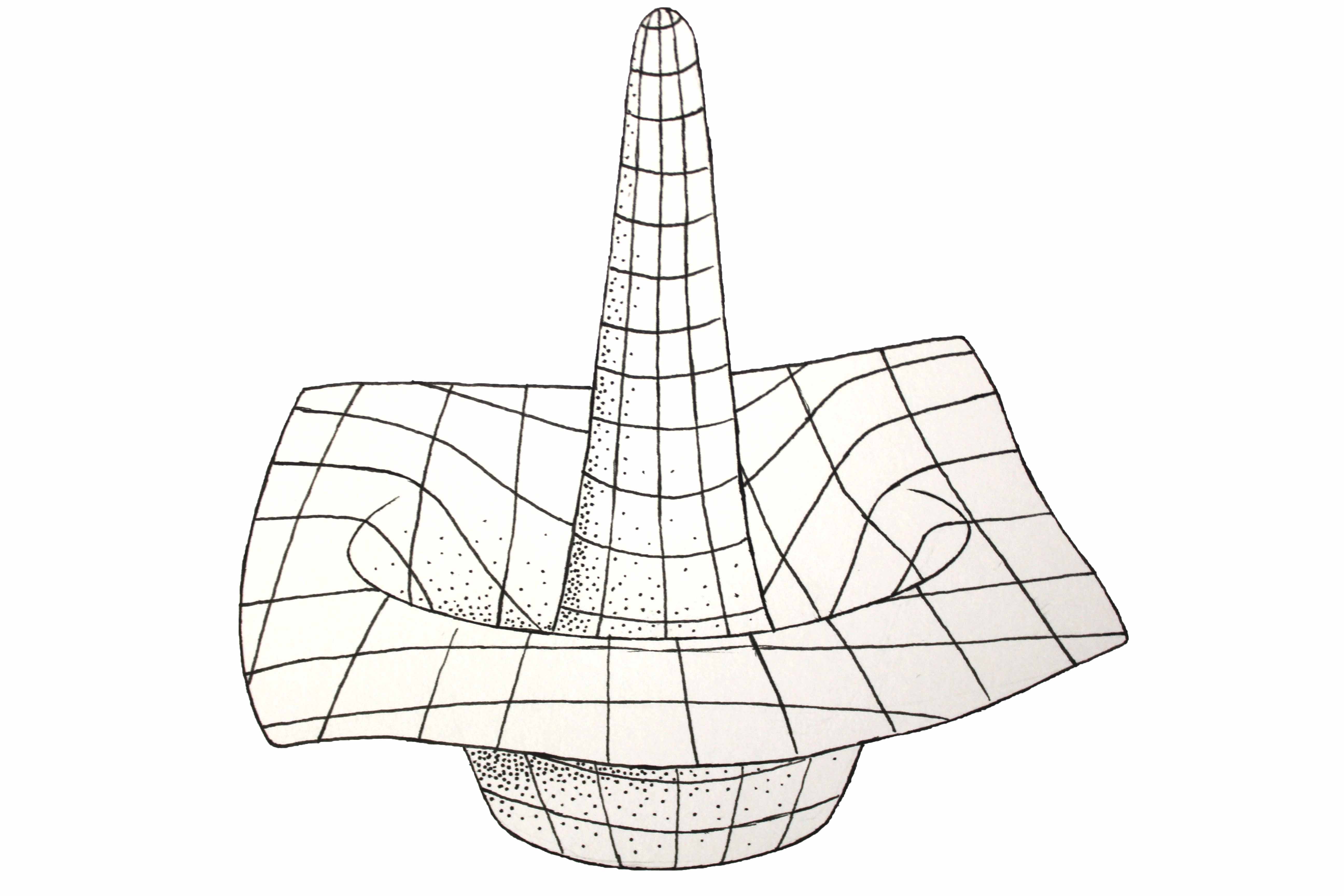}	
	\end{center}
	\caption{Diagram showing the qualitative form of the interaction potential.} \label{Fig:potential}
\end{figure}

This interaction has $\tilde{V}_{l\geq2}=0$. Near the Fermi surface, the remaining components verify $|\tilde{V}_1|\gg|\tilde{V}_0|$ so that the potential approximately behaves as (\ref{eq:foucomp}). 
This interaction is the distributional limit of potentials of the form shown in Figure~\ref{Fig:potential}. These potentials exhibit a repulsive hard core and an attractive tail (precisely the type of interaction between $^3$He atoms). As it is not possible to construct a translation-invariant interaction potential with only $\tilde{V}_1\neq0$ (it would fail to be invariant under constant shifts in momentum space), (\ref{dpotential}) is the best approximation to an interaction of the form (\ref{eq:foucomp}) one can find.

For properties involving long wavelengths compared with the interparticle distance, the model potential~(\ref{dpotential}) will be perfectly appropriate as representative of an entire microscopic class. Taking this potential, the Grand Canonical pairing Hamiltonian finally reads
\begin{eqnarray}
H_\pairing -\mu N:=&
\sum_{\bm{p} \alpha} 
\left({p^2 \over 2m^*} -\mu \right)
{a}_{\bm{p}\alpha}^\dagger {a}_{\bm{p}\alpha}
\nonumber\\
&-\frac{g}{2p_\fermi^2} \sum_{\bm{p} \bm{p}' \alpha \beta} (\bm{p}'\cdot \p) {a}_{-\bm{p}' \beta}^\dagger {a}_{\bm{p}' \alpha}^\dagger
{a}_{\bm{p} \alpha} {a}_{-\bm{p} \beta}.
\label{eq:hampairing}
\end{eqnarray}
This is the system we will work with in the next subsections.

%

\subsection{Condensation and order parameters \label{sec:order}}

The model interaction described in the previous subsection is called a $p$-wave spin-triplet pairing interaction. Below a critical temperature it enforces the formation of anisotropic Cooper pairs (as opposed to the isotropy of the Cooper pairs in classical superconductivity). The spatial anisotropy of these pairs is associated with the fact that they possess angular momentum. Given the antisymmetric structure of the orbital part of the wave function, its spin structure has to be symmetric and thus  belongs to the triplet space of the spin product. These pairs condense acquiring a macroscopic occupation. The macroscopic wave function or order parameter associated with the condensed pairs will be
\begin{eqnarray}
\mathbf{\Psi}_{\alpha\beta}:=\frac{g}{p_\fermi}\bigg\langle \sum_{\p} \bm{p}\,{a}_{\bm{p} \alpha} {a}_{-\bm{p} \beta}\bigg \rangle~.\label{eq:orderpar}
\end{eqnarray}
As a consequence of the spin-dependence and of the dominance of anisotropic $p$-wave interaction, this order parameter is not a scalar (as in the case of classical superconductivity or Bose-Einstein condensation) but a matrix, with spin indices $\alpha$, $\beta$. There is also an implicit orbital index $i$ because of the $\p$-dependence of $\mathbf{\Psi}_{\alpha\beta}$.
The normal-liquid phase has as symmetry group $\mbox{SO}(3)_{_{\rm{ L}}} \times \mbox{SO}(3)_{_{\rm{S}}} \times \mbox{U}(1)$, {\it i.e.} independent rotations of the coordinate and spin spaces plus a phase-invariance symmetry associated with the conservation of the number of atoms. Pair condensation amounts to the spontaneous (partial) breaking of this symmetry. The order parameters appearing in this $p$-wave spin-triplet condensation are symmetric in the spin indices and therefore can always be written as: 
\begin{eqnarray}
\Psi^i_{\alpha\beta}=i (\sigma_a \sigma_2)_{\alpha\beta} \, d^{ai}~,
\label{eq:orderparameter} 
\end{eqnarray}
where $\sigma_a$ are Pauli matrices and $d^{ai}$ is in general a complex vector in both spin  and position space. The set $\{\sigma_a \sigma_2\}_{a=1,2,3}$ forms a basis for all $2\times 2$ symmetric matrices. The imaginary prefactor $i$ is introduced by convention to make this a real matrix basis and put all the complexity into the vector $d^{ai}$.

Depending on the details of the interaction, the order parameter can acquire different structures. The precise form of the order parameter is obtained by a minimization principle. In the microscopic theory the quantity to be minimised is the expectation value of the Hamiltonian~(\ref{eq:hampairing}) in the Fock vacuum state. One can alternatively use a minimization within Ginzburg-Landau theory, which is a special case of the phenomenological Landau-Lifshitz theory of second-order phase transitions. In this approach one has to minimize the free-energy functional of the order parameter. The order parameter~(\ref{eq:orderparameter}) is zero above a certain critical temperature $T_\critical $ but takes a finite value for $T<T_\critical $. The thermodynamic potential of interest in the experimental situation (constant temperature $T$ and volume $V$) is the Helmholtz free energy. If we suppose that near $T_\critical $ the free energy is analytic in the order parameter and  obeys the symmetries of the microscopic Hamiltonian, then one can write a Taylor expansion near the critical temperature. The symmetries of the interaction dictate the type of terms that can appear in this expansion. The precise values of the coefficients in Ginzburg-Landau theory depend on the microscopic theory and can be derived from  it, for example in the BCS theory of superconductivity~\cite{Gorkov1959}.

\ref{sec:GL} provides some details regarding  the minimization procedure  in the case in which one neglects the spin-spin interactions. Under certain conditions, four solutions are found to the minimization problem. The BW (Balian-Werthamer) and ABM (Anderson-Brinkman-Morel) states are associated by confrontation with experiments with the superfluid phases B and A, respectively. The other two states are the so-called planar and polar states. The planar state and the ABM state are topologically characterised by the presence of Fermi points. The BW state is fully gapped, while the polar state has a Fermi manifold of dimension 1. As we will see, Fermi points give rise to relativistic low-energy excitations. It is easy to understand why this is the case: near these points, the dispersion relation of quasiparticles is linear to leading order, and is three-dimensional, unlike in the case of Fermi manifolds of higher dimension~\cite{Horava2005}.

For our purposes, we are specially interested in the planar state; we will see clearly why in the next subsections. Its order parameter is  
\begin{eqnarray}
d_\textrm{\scriptsize planar}^{a i}(T):=\Delta(T)(\hat{s}^{a}\hat{m}^{i}+\hat{s}'^{a}\hat{n}^{i}),\label{eq:plaord}
\end{eqnarray}
where $\hat{\m}$, $\hat{\n}$ are unit vectors in position space and $\hat{\bm s}$ and $\hat{\bm s}'$ unit vectors in spin space subject to the orthogonality conditions $\hat{\m}\cdot \hat{\n}=0$, $\hat{\bm s} \cdot \hat{\bm s}'=0$. In this expression, the scalar function $\Delta(T)$ is the gap parameter which contains the dependence of the order parameters on the temperature $T$ and the interaction constant $g$. At zero temperature its value is approximately $\Delta_0:=\Delta(0)\simeq k_\boltzmann T_\critical $, where $T_\critical $ is the critical temperature.

The planar state has not yet been observed in nature among the superfluid phases of $^3$He. If one neglects the dipole-dipole interactions, then this state is never the lowest energy state of the system. However, when taking into account these interactions, which in $^3$He are rather feeble, this state should be the global minimum in a narrow temperature band in phase space~(see for example~\cite{dePrato2004}). Here we are not considering real $^3$He but a system constructed with atoms adapted to our needs. Thus we will just assume that there exist some additional atom-atom interactions beyond the Lennard-Jones potentials such that they select the planar state as the natural vacuum.

It is sometimes instructive to have in mind the other well-known states of this system: the ABM and BW states. Their order parameters are respectively
\begin{eqnarray}
d_\textsc{\footnotesize abm}^{a i}(T):=\Delta(T)\hat{s}^a(\hat{m}^i+i\hat{n}^i),\label{eq:abmord}\\
d_\textsc{\footnotesize bw}^{a i}(T):=\Delta(T) \delta^{ai}.
\label{eq:bword}
\end{eqnarray}
Here there is also an orthogonality condition $\hat{\m}\cdot \hat{\n}=0$.

Before closing this subsection let us comment that, within the interpretation of a strongly interacting system of atoms, the realization of any of these condensed phases takes us beyond the strict limits of applicability of Landau's Fermi-liquid hypothesis. The Fermi surface of the free system has been deformed so strongly that it no longer survives. It has been either completely eliminated (BW state) or reduced to just some points (planar and ABM states). However, it is remarkable that a weak-interaction model of quasiatoms is able to describe correctly the condensation and low-energy excitation of these systems. For the interpretation in which one directly starts from a weakly interacting system of atoms, the previous comment is irrelevant: in this case the weakly interacting theory is already the very microscopic theory.

\subsection{Low-energy quasiparticle excitations}
\label{subsec:low-energy}

In this subsection we will analyse  the form of the quasiparticle excitations living right above the vacuum of the planar state. These are new types of quasiparticles, specific combinations of the atoms and holes
of Landau's theory. We will eventually call them Bogoliubov quasiparticles.
 
Once the system has settled to a condensed state, the pairing interaction can be expanded up to quadratic order in perturbations around the condensed state (the so-called Gor'kov factorization \cite{Gorkov1958}). It is easy to see that the resulting quadratic Hamiltonian reads
\begin{eqnarray}
H_\pairing -\mu N:=&
\sum_{\bm{p} \alpha} 
M(\bm{p})
{a}_{\bm{p}\alpha}^\dagger {a}_{\bm{p}\alpha}
\nonumber\\
&+\frac{1}{2p_\fermi}\sum_{\bm{p} \alpha \beta} \bm{p} \cdot\mathbf{\Psi}_{\alpha\beta} {a}_{-\bm{p} \beta}^\dagger {a}_{\bm{p} \alpha}^\dagger
+\frac{1}{2p_\fermi}\sum_{\bm{p} \alpha \beta} \bm{p} \cdot\mathbf{\Psi}^*_{\beta\alpha} {a}_{\bm{p} \alpha} {a}_{-\bm{p} \beta}~,
\label{eq:hampairing3}
\end{eqnarray}
where we have defined $M(\bm{p})=p^2/(2m^*) -\mu$. Consider now the order parameter to be a homogeneous planar state characterised by the vectors $\hat{\bm{s}},\hat{\bm{s}}',\hat{\bm{m}},\hat{\bm{n}}$. Let us choose a system of coordinates adapted to the pairs-spin-space Cartesian trihedral 
\begin{equation}
\hat{\bm{x}}=\hat{\bm{s}},\qquad
\hat{\bm{y}}=\hat{\bm{s}}',\qquad
\hat{\bm{z}}=\hat{\bm{s}} \times \hat{\bm{s}}'.
\end{equation}
Calculating the commutator between quasiparticle operators and $H_\pairing -\mu N$ shows that the evolution equations of quasiparticle operators particularised to the planar order parameter (\ref{eq:plaord}) are
\begin{eqnarray}
i\dot{a}_{\bm{p}\uparrow}=M(\bm{p})a_{\bm{p}\uparrow}-c_\bot\bm{p}\cdot(\hat{\bm{m}}-i\hat{\bm{n}})a^\dagger_{-\bm{p}\uparrow},
\label{evoup}
\\
i\dot{a}_{\bm{p}\downarrow}=M(\bm{p})a_{\bm{p}\downarrow}+c_\bot\bm{p}\cdot(\hat{\bm{m}}+i\hat{\bm{n}})a^\dagger_{-\bm{p}\downarrow}.
\label{evodown}
\end{eqnarray}
Here we have introduced the parameter $c_\bot:=\Delta_0/p_\fermi$ with dimensions of velocity. The evolution of the two spin populations is decoupled. This property permits us to consider the spin populations separately, simplifying the treatment.

Let us first anticipate the appearance of Fermi points in this vacuum state. Acting with the operator $i\partial_t$ on (\ref{evoup}), one finds that the dependence on $a^\dagger_{-\bm{p}\uparrow}$ vanishes, and one can write
\begin{eqnarray}
(i\partial_t)^2 {a}_{\bm{p}\uparrow}&=\left\{ M^2(\bm{p})+c_\bot^2 [(\bm{p}\cdot(\hat{\bm{m}}-i\hat{\bm{n}})  (\bm{p}\cdot(\hat{\bm{m}}+i\hat{\bm{n}})] \right\} a_{\bm{p}\uparrow} \nonumber\\
&=\left\{ 
M^2(\bm{p})+
c_\bot^2 (\bm{p}\times \hat{\bm{l}})^2
\right\} a_{\bm{p}\uparrow},
\end{eqnarray}
where $\hat{\bm{l}}:=\hat{\bm{m}}\times\hat{\bm{n}}$. The eigenvalues of the evolution operator vanish only in the so-called Fermi points in momentum space\footnote[1]{Also often called Weyl points. We use the term ``Fermi point'' in accordance with \cite{Volovik2003}. ``Fermi point'' can be understood as the generic term for topological point nodes, which includes Weyl points when the underlying manifold is 3+1 dimensional, and Dirac points for 2+1 dimensions. $^3$He-A is then an example of the Weyl category of Fermi points.},
\begin{equation}
\bm{p}_{\fermi,\pm} = \pm p_\fermi\hat{\bm{l}}~,
\end{equation}
as these eigenvalues are given by the square root of
\begin{eqnarray}
M^2(\bm{p})+c_\bot^2 (\bm{p}\times \hat{\bm{l}})^2.
\end{eqnarray}
We can now see that the dispersion relation is relativistic near these points in momentum space. Considering a plane wave with momentum $\p=qp_\fermi \hat{\bm{l}}+\mathfrak{p}$ with $q=\pm 1$ and $\mathfrak{p}$ a small deviation with respect to the corresponding Fermi point, we obtain the frequency 
\begin{equation}
\omega^2 = c_\|^2 \mathfrak{p}_l^2 + c_\bot^2 (\mathfrak{p}_m^2 + \mathfrak{p}_n^2),\label{eq:disp}
\end{equation}
where $c_\| = p_\fermi/m^*$ (recall that $p_F=\sqrt{2m^*\mu}$). We use the subindices $m,n,l$ to denote projections on the pairs Cartesian trihedral $\hat{\bm{m}},\hat{\bm{n}},\hat{\bm{l}}$, {\it i.e.}  $\mathfrak{p}_m=\mathfrak{p}\cdot \hat{\bm{m}}$, etc. This linear dispersion relation is valid below the energy scale
\begin{equation}
E_\lorentz:= m^*c_\bot^2.\label{eq:lorentz}
\end{equation}
A look at Eq. (\ref{eq:disp}) reveals that the parameters $c_\|$ and $c_\bot$ correspond to the propagation velocity    of low-energy quasiparticles in the directions parallel and perpendicular to the anisotropy axis, respectively.

Now we want to diagonalize the Hamiltonian but concentrating on two regions in momentum space surrounding the two Fermi points (these regions contain the real low-energy excitations of the system). In other words, we shall find new annihilation operators $\alpha_{\bm p\alpha q}$  over which the action of the Hamiltonian is  diagonal. As labels for these operators we use the deviation $\mathfrak{p}$ with respect to any of the Fermi points, $\bm{p}=\pm p_{\fermi}\hat{\bm{l}}+\mathfrak{p}$, the spin index $\alpha$, and   a subscript $q=u,d$ indicating the Fermi point near which it is localised (in momentum space): the $u$ Fermi point ($+p_{\fermi}\hat{\bm{l}}$) or the $d$ Fermi point ($-p_{\fermi}\hat{\bm{l}}$).
In terms of the label $\mathfrak{p}$ this leads to an apparent doubling of the degrees of freedom. Alternatively, one can work directly with operators $a_{\mathfrak{p}\alpha q}$ (the only difference between these two sets of operators is a linear change of basis) defined as
\begin{equation}
a_{\mathfrak{p} \uparrow u}:=a_{p_\fermi\hat{\bm{l}}+\mathfrak{p},\uparrow},\qquad a_{\mathfrak{p}\uparrow d}:=a_{-p_\fermi\hat{\bm{l}}+\mathfrak{p},\uparrow}.
\end{equation}
Focusing first on the $\uparrow$ spin projection, we can write the corresponding equations of motion as
\begin{eqnarray}
i\dot{a}_{\mathfrak{p}\uparrow u} =
c_\| \mathfrak{p}_l a_{\mathfrak{p}\uparrow u}
-c_\bot (\mathfrak{p}_m -i \mathfrak{p}_n) a_{-\mathfrak{p}\uparrow d}^\dagger~,
\nonumber\\
i\dot{a}^\dagger_{\mathfrak{-p}\uparrow d} =
-c_\| \mathfrak{p}_l a_{-\mathfrak{p}\uparrow d}^\dagger
-c_\bot (\mathfrak{p}_m +i \mathfrak{p}_n) a_{\mathfrak{p}\uparrow u}~,
\end{eqnarray}
remembering that the equality sign is strictly speaking an approximately-equal sign, and that it is not valid for momenta too far from the Fermi point. The two previous equations can be written in a compact manner as 
\begin{equation}
i\partial_t \chi_{\mathfrak{p}\uparrow} = \mathscr{H}_{\mathfrak{p\uparrow}}\chi_{\mathfrak{p}\uparrow}~,
\qquad
\chi_{\mathfrak{p}\uparrow}=\left(\begin{array}{c}a_{\mathfrak{p}\uparrow u}\\ a^\dagger_{-\mathfrak{p}\uparrow d}\\\end{array}\right)~,\label{eq:ngup}
\end{equation}
with
\begin{equation}
\mathscr{H}_{\mathfrak{p}\uparrow}:=c_\| \mathfrak{p}_l\sigma_3-c_\bot \mathfrak{p}_m \sigma_1 -c_\bot \mathfrak{p}_n \sigma_2.
\end{equation}
This is a linear spinor equation for a Weyl spinor with helicity $+$ (calculated as the product of the three factors $\pm 1$ that appear in front of the Pauli matrices). Before continuing let us notice that the evolution equations for all the $a_{\bm{p}\uparrow}$ in eq.~(\ref{evoup}) are not linear in the complex plane due to the presence of complex conjugate terms. However, they have a different quasilinear symmetry. The system is invariant if one multiplies the $a_{\bm{p}\uparrow}$ with $\bm{p}$ in the $u$ hemisphere by a complex constant $c$ and those in the $d$ hemisphere by its complex conjugate $c^*$. This symmetry has allowed us to write a linear equation for the previous spinor $ \chi_{\mathfrak{p}\uparrow}$. This spinor contains information about both Fermi points.

The same arguments can be applied to the $\downarrow$ projection of the real (atomic) spin of Landau's quasiparticles to obtain
\begin{equation}
\chi_{\mathfrak{p}\downarrow}:=\left(\begin{array}{c}a_{\mathfrak{p} \downarrow u}\\
a^\dagger_{-\mathfrak{p}\downarrow d}
\end{array}\right)\label{eq:ngdown},
\end{equation}
and Hamiltonian operator
\begin{equation}
\mathscr{H}_{\mathfrak{p}\downarrow}:=
c_\| \mathfrak{p}_l\sigma_3+c_\bot \mathfrak{p}_m\sigma_1-c_\bot \mathfrak{p}_n\sigma_2,
\end{equation}
with helicity $-1$ in this case. Notice that the only difference between the two spin projections is a sign accompanying $\hat{\bm{m}}$ in the order parameter [see also Eqs.~(\ref{evoup},\ref{evodown})], which translates into a different helicity in the low-energy corner. For this reason the atomic spin projection index can be thought of as a helicity index for the low-energy Bogoliubov excitations. We will explicitly check this later.

As a final step one can arrange the two spinors to form a bispinor that obeys the following evolution equation:
\begin{equation}
i\partial_t 
\left(\begin{array}{c}\chi_{\mathfrak{p}\uparrow}\\\chi_{\mathfrak{p}\downarrow}\\\end{array}\right)
=e_{\ b}^a Y^b \mathfrak{p}_a ~
\left(\begin{array}{c}\chi_{\mathfrak{p}\uparrow}\\\chi_{\mathfrak{p}\downarrow}\\\end{array}\right),\label{eq:plamat0}
\end{equation}
with $a,b=1,2,3$ and
\begin{equation}
Y^1=\left(\begin{array}{cc}-\sigma_1&0\\0&\sigma_1\\\end{array}\right),\quad Y^2=\left(\begin{array}{cc}-\sigma_2&0\\0&-\sigma_2\\\end{array}\right),\quad Y^3=\left(\begin{array}{cc}\sigma_3&0\\0&\sigma_3\\\end{array}\right).\label{eq:plamat}
\end{equation}
The only non-zero components of $e^b_{\ a}$ are
\begin{equation}
e^1_{\ 1}:=c_\bot,\qquad e^2_{\ 2}:=c_\bot,\qquad e^3_{\ 3}:=c_\|.\label{eq:tetr}
\end{equation}
Now we can try to find a matrix $X$ such that the set $\{X,XY^1,XY^2,XY^3\}$ is a representation of the Dirac matrices. Taking into account that the matrices (\ref{eq:plamat}) verify the properties
\begin{equation}
(Y^a)^2=I_4,\qquad \{Y^a,Y^b\}=2\delta^{ab}I_4,
\end{equation}
(with $I_4$ the $4\times4$ identity), which follow directly from the properties of the Pauli matrices, such a matrix $X$ must verify
\begin{equation}
X^2=I_4,\qquad \{X,Y^a\}=0.
\end{equation}
One can check that a  solution to these equations is given by
\begin{equation}
X=\left(\begin{array}{cc}0&\sigma_1\\\sigma_1&0\\\end{array}\right).
\end{equation}
The corresponding representation of the Dirac matrices is:
\begin{eqnarray}
\gamma^0=\left(\begin{array}{cc}0&\sigma_1\\\sigma_1&0\\\end{array}\right),\qquad \gamma^1=\left(\begin{array}{cc}0&I_2\\-I_2&0\\\end{array}\right),\nonumber\\
\gamma^2=\left(\begin{array}{cc}0&-i\sigma_3\\-i\sigma_3&0\\\end{array}\right),\qquad \gamma^3=\left(\begin{array}{cc}0&-i\sigma_2\\-i\sigma_2&0\\\end{array}\right).
\end{eqnarray}
So we can conclude that the low-energy excitations of the planar phase are massless Dirac spinors in Minkowski spacetime, thus satisfying the evolution equation
\begin{equation}
e^\mu_{\ I}\gamma^I\bar \mathfrak{p}_\mu \psi_{\mathfrak{p}}=0,\qquad \psi_{\mathfrak{p}}:=\left(\begin{array}{c}\chi_{\mathfrak{p}\uparrow}\\\chi_{\mathfrak{p}\downarrow}\\\end{array}\right),
\label{eq:dirace0}
\end{equation}
where we have taken the Fourier time transform and defined $\bar\mathfrak{p}^\mu:=(\omega,\mathfrak{p})$. The components of the tetrad $e^\mu_{\ I}$, $\mu=1,2,3,4$, $I=1,2,3,4$ are given by (\ref{eq:tetr}) complemented by
\begin{equation}
e^0_{\ 0}:=1.
\end{equation}
Spacetime is therefore Minkowskian because the tetrad field has been taken to be constant. The constant velocities $c_\|$ and $c_\bot$ can be absorbed into a rescaling of the coordinates. This laboratory anisotropy would in any case be unobservable for low-energy ``internal'' observers (see next section and \cite{Barcelo2008}).

The occurrence of four components in the low-energy fermionic object $\psi$ (whose Fourier components are defined in (\ref{eq:dirace0})) is tied up to the existence of two degrees of freedom, one for each Fermi point, for each projection of the spin. The spin projection must be considered as the helicity eigenvalue in the low-energy description: let us evaluate the chirality operator in this representation,
\begin{eqnarray}
\gamma^5:=i\gamma^0\gamma^1\gamma^2\gamma^3=\left(\begin{array}{cc}I_2&0\\0&-I_2\\\end{array}\right).
\end{eqnarray}
That is, $\chi_\uparrow$ and $\chi_\downarrow$ have a well-defined chirality, as
\begin{equation}
\frac{1+\gamma^5}{2}\psi=\psi_\uparrow =
\left(\begin{array}{c} \chi_\uparrow \\ 0\end{array}\right),
\qquad 
\frac{1-\gamma^5}{2}\psi=\psi_\downarrow =
\left(\begin{array}{c}0 \\ \chi_\downarrow \end{array}\right).
\end{equation}

Summarizing, the natural low-energy excitations that show up in this model are massless Dirac fermions. To reproduce electrodynamics of electrons and positrons, one would need to generate some small mass gap for the excitations. This might require a more complicated system, with for example some Yukawa couplings, and is beyond the analysis carried out here.

\subsection{Internal low-energy observers and the emergence of charge}   

 We have seen that the concept of atomic spin acquires a different meaning in the low-energy corner in which the fermionic quasiparticles are described by Dirac's theory: they play the role of a charge label. The Dirac equation in (\ref{eq:dirace0}) is invariant under a $\mbox{U}(1)$ transformation of $\psi$ (in fact it is invariant under transformations of $\psi_\uparrow$ and $\psi_\downarrow$ separately). The corresponding conserved charge is
\begin{equation}
Q:=Q_\uparrow+Q_\downarrow=N_u-N_d,\label{eq:charge1}
\end{equation}
where the operators $N_u$ and $N_d$ represent the number of excitations associated with the positive Fermi point (with positive $q=1$ charge) and with the negative Fermi point (with negative $q=-1$ charge), respectively. 
As we are going to explain further, a notion of charge has emerged in the low-energy theory owing to the duplicity of Fermi points. In fact we will see later that when coupling these Dirac  quasiparticles to an effective electromagnetic field, this charge plays the role of an electric charge.

For an external observer (or laboratory observer), charge conservation is nothing more than momentum conservation. Imagine a scattering process involving two quasiparticles from {\em the same Fermi point}. Momentum conservation only tells us that 
\begin{equation}
\p_1 +\p_2 = 2qp_\fermi\hat{\bm{l}}+\mathfrak{p}_1+\mathfrak{p}_2=\bm{p}_3+\bm{p}_4.
\end{equation}
As the products of the scattering event must be quasiparticles, the solutions of this equation must be of the form
\begin{equation}
\bm{p}_3=qp_\fermi\hat{\bm{l}}+\mathfrak{p}_3,\qquad \bm{p}_4=qp_\fermi\hat{\bm{l}}+\mathfrak{p}_4.
\end{equation}
The momentum conservation condition is thus equivalent to the conservation of the deviations $\mathfrak{p}$:
\begin{equation}
\mathfrak{p}_1+\mathfrak{p}_2=\mathfrak{p}_3+\mathfrak{p}_4.
\end{equation}
If the scattering is instead between quasiparticles from {\em different Fermi points} the conservation of momentum reads
\begin{equation}
\p_1 +\p_2 = \mathfrak{p}_1+\mathfrak{p}_2=\bm{p}_3+\bm{p}_4,
\end{equation}
which implies that
\begin{equation}
\bm{p}_3=qp_\fermi\hat{\bm{l}}+\mathfrak{p}_3,\qquad \bm{p}_4=-qp_\fermi\hat{\bm{l}}+\mathfrak{p}_4.
\end{equation}
The resulting quasiparticles have to live in different Fermi points.
Again, the momentum conservation condition is then equivalent to the conservation of the deviations $\mathfrak{p}$,
\begin{equation}
\mathfrak{p}_1+\mathfrak{p}_2=\mathfrak{p}_3+\mathfrak{p}_4.
\end{equation}
In both cases, the energy-conservation condition is equivalent at low-energies to the usual massless relativistic condition in the deviations $\mathfrak{p}$, so it does not distinguish between the two types of processes.

\begin{figure}
	\begin{center}
		\includegraphics[width=.9\columnwidth]{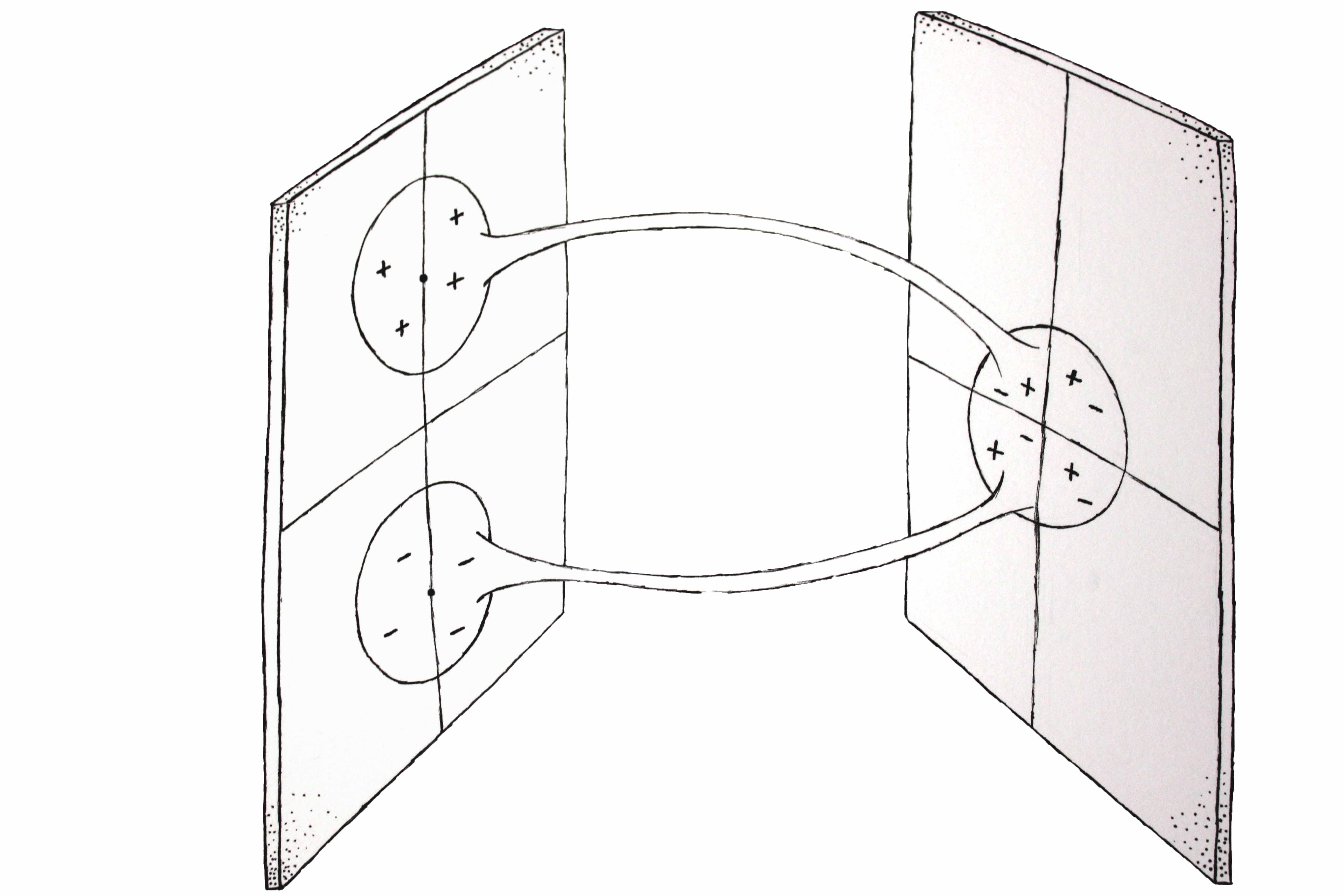}	
	\end{center}
	\caption{Diagram explaining the effective duplication of the degrees of freedom. The left-hand side shows the quasiparticles living near the two Fermi points. The right-hand side shows the effective low-energy description for an internal observer. An internal observer is insensitive to the origin (upper or lower Fermi point) of the quasiparticles in terms of momentum, but he sees them as having opposite charges.} \label{Fig:duplication}
\end{figure}

However, in the low-energy description, the conserved charge is associated with an intrinsic property of the Dirac field and its conservation has nothing to do, from this point of view, with momentum conservation. The effective duplication of degrees of freedom in the low-energy theory with respect to the degrees of freedom in the initial quasiparticle field is just apparent. We can understand it by looking at Figure~\ref{Fig:duplication}.   

In this paper we are interested in constructing an effective low-energy world that cannot be distinguished operationally from the world of electrodynamics. We therefore have to discuss the important meaning of internal observer. 
In relation with the emergence of charge, here we can already discuss the difference between two types of potential internal observers. One can be called an internal Fermi-point observer, the other an internal low-energy observer.

An internal Fermi-point observer is an observer living in one specific Fermi point $q$. We can associate a momentum $qp_\fermi\hat{\bm{l}}$ to that observer. He will see the momentum region around him as a low-energy world full of spinor waves (these will not be Weyl spinors but specific superpositions of them). His world would have half the degrees of freedom compared to the Dirac bispinors. In addition, this observer will see quasiparticles coming from the other Fermi point, which will have a tremendous relative momentum $2qp_\fermi$, although they will have low energies. To obtain a standard electrodynamic world for these kind of observers our model lacks two ingredients: i) the quasiparticles from different Fermi points should not interact with each other (in the model we are discussing they do); ii) one should duplicate in some way the number of degrees of freedom associated with that Fermi point (maybe producing some fragmented Fock state condensation).

An internal low-energy observer on the other hand is an observer who sees {\em all} the low-energy excitations. It is reasonable that they will use as a natural momentum label the deviations $\mathfrak{p}$. These momenta can properly describe the scattering events between all the quasiparticles as long as these observers confer an additional property to these quasiparticles, which is conserved in the interaction process. This property is charge, even though for the external observers this is nothing but the difference between the quasiparticle number around both Fermi points, as we have seen. 

Let us discuss one final point regarding the nature of the low-energy excitations of the system (the previously described spin waves or Dirac quasiparticles). Consider a spin wave with exactly the Fermi point momentum
\begin{equation}
e^{ip_\fermi \hat{\bm{l}} \cdot \bm{x}/\hbar}~.
\end{equation}
We have seen that this oscillation pattern carries no energy. This oscillation is stationary so it cannot carry momentum either. This oscillation pattern is in reality part of the vacuum state. If the momentum of these spin waves has a small departure $\mathfrak{p}$ from the Fermi point momentum, then we have seen that they do carry an energy $E \simeq c|\mathfrak{p}|$. As we have explained, the effective spacetime is Minkowskian, so the anisotropic velocities in the laboratory will not have any operational meaning for low-energy internal observers. We will just define a single constant $c$ relating their space and time dimensions. The momentum carried by one such  wave is precisely $\mathfrak{p}$. Its direction marks the direction of the propagation of the spin wave. Its modulus can be seen as derived from $E/c$. Therefore, the momentum $p_\fermi \hat{\bm{l}}+\mathfrak{p}$ is not the real momentum carried by the spin wave, relevant to experiments measuring impulse transfers within the liquid. The real momentum of the spin wave is  $\mathfrak{p}$.

A momentum $|\mathfrak{p}|$ has an associated wavelength $\lambda=2\pi\hbar/|\mathfrak{p}|$. These wavelengths, which are much larger than the mean interparticle distance $\lambda_{_{\rm{I}}}=2\pi\hbar/p_{\fermi}$, are the actual ``observable'' wavelengths of the spin waves in the liquid. In the case of a classical (large amplitude) spin wave this ``observability'' will match our intuitive sense of observation of a wave.

\subsection{Textures and electromagnetic fields} \label{sec:text}  

Let us analyse now what happens in the case in which the order parameter, instead of being completely homogeneous, contains a perturbation in position space with respect to the situations of the previous subsection. Such inhomogeneities in this kind of order parameters are typically called textures. These variations develop over a scale which is large compared to the effective size of a Cooper pair, or healing length, whose zero-temperature limit is
\begin{equation}
\xi_0:=\frac{\hbar}{p_\fermi}\frac{c_\|}{c_\bot}\simeq\frac{\hbar p_\fermi}{ 
 m^*k_\boltzmann T_\critical }\label{eq:sizecp}.
\end{equation}
In other words, the Fourier modes of the variations of the order parameter have wave numbers given by
\begin{equation}
k\ll k_\textrm{\scriptsize max}:=\frac{2\pi}{\xi_0}.\label{eq:kmax}
\end{equation}
This means that all wavelength variations must be much larger than the healing length (\ref{eq:sizecp}), since shorter wavelength variations are not consistent with the very existence of a local order parameter. An equivalent restriction applies to the rapidity of temporal variations of the order parameter. If we define a natural time scale as $t_0:=\xi_0/c_\perp$, then for consistency we have to be sure that the temporal variations of the order parameter are slower than this time scale.

The textures we will consider are of two kinds. The first one is usually called orbital texture \cite{Leggett2006} and is given by the bending of the direction $\hat{\bm{l}}$ of the angular momentum of the pairs. They amount to two degrees of freedom. In addition to this, the planar-phase order parameter has, in general, the possibility of rotating around the angular momentum axis. This leads to one additional degree of freedom. In simple situations, when only this rotation is present, the additional degree of freedom is just a phase from which one can define a superfluid velocity and momentum as
\begin{equation}
\bm{v}_\superfluid (\bm{x}):= {1 \over 2m^*} \bfnabla \phi~,~~~~ \bm{p}_\superfluid (\bm{x}):=m^*\bm{v}_\superfluid (\bm{x})~.
\end{equation}
In more complicated situations the superfluid velocity need not be irrotational (see~\cite{Mermin1976}), but the important thing is that, in any case, the superfluid velocity contributes with one single additional degree of freedom to the physics of the system, on top of the two degrees of freedom of $\delta \hat{\bm{l}}$.  
In the simplest case in which $\bm{p}_\superfluid $ is a constant vector
we can again work in momentum space to analyse the form of the low-energy excitations. 

The selection of a specific inertial frame in which the Fermi fluid is at rest can be seen as a peculiar example of spontaneously broken symmetry. Two relatively moving states have different energies with respect to a third inertial observer ({\it e.g.} the laboratory observer). However, if there were no interaction at all between the fluid and external objects in a particular frame, there would be no physical reason to select one specific uniform fluid velocity rather than another. In practice, the tiny interactions between the fluid and some specific inertial environment (typically the laboratory environment/frame) induces this very frame as the rest frame of the fluid. Then, the condensed vacuum state incorporates this same frame selection: the pairs are at rest with respect to this specific frame. 

In what follows we take the operational view that a specific frame with a constant velocity $\bm{v}_\superfluid $ with respect to the laboratory has been selected, regardless of the origin of this selection. 
This means that the pairing has occurred between $\bm{p} + \bm{p}_\superfluid $ and $-\bm{p} + \bm{p}_\superfluid $ atoms. This implies that 
eqs. (\ref{evoup},\ref{evodown}) should now be written as  
\begin{eqnarray}
i\dot{a}_{\bm{p}+\bm{p}_\superfluid ,\uparrow}=[M(\bm{p})+\bm{p}\cdot\bm{v}_\superfluid] a_{\bm{p}+\bm{p}_\superfluid ,\uparrow}
-c_\bot \bm{p}\cdot(\hat{\bm{m}}-i\hat{\bm{n}})
a^\dagger_{-\bm{p}+\bm{p}_\superfluid ,\uparrow}~,
\label{evoup-super}
\\
i\dot{a}_{\bm{p}+\bm{p}_\superfluid ,\downarrow}=[M(\bm{p})+\bm{p}\cdot\bm{v}_\superfluid] a_{\bm{p}+\bm{p}_\superfluid ,\downarrow}+c_\bot \bm{p}\cdot(\hat{\bm{m}}+i\hat{\bm{n}})a^\dagger_{-\bm{p}+\bm{p}_\superfluid ,\downarrow}~.
\label{evodown-super}
\end{eqnarray}
To reach these equations one needs to perform an active Galilean transformation under which any momentum label is shifted by $+\bm{p}_\superfluid $, and take into account the transformation laws for the different objects appearing in the evolution operator.  This is best understood from the Galilean transformation of the Grand Canonical Hamiltonian~(\ref{eq:mbhammom}). Recall that the potential term is invariant under such transformation and that the kinetic term acquires two extra terms: a Doppler contribution $ \bm{p}\cdot\bm{v}_\superfluid $ and a global shift $ {\bm{p}_\superfluid ^2}/({2m^*})$, which can be absorbed
in the chemical potential for the moving system~\cite{Ho1980},
\begin{equation}
\bar{\mu}=\mu +\frac{\bm{p}_\superfluid ^2}{2m^*}~;
\end{equation}
\begin{equation}
M(\bm{p})+\bm{p}\cdot\bm{v}_\superfluid = 
{(\bm{p}+\bm{p}_\superfluid)^2 \over 2 m^*} - \bar{\mu}~.
\end{equation}
If we take into account that the pairing channel is given by   $\p_1+\p_2=\p_3+\p_4=0$,  we immediately reach the conclusion that the transformed pairing Hamiltonian is precisely (\ref{eq:hampairing}) with a Doppler shift and with all the labels shifted by $\bm p_\superfluid$. Finally, the order parameter is just book-keeping the statistics of the pairs and hence depends only on the relative momenta of the members of the pairs and not on their global motion, which means that the order parameter is unchanged by the Galilean transformation.

As an alternative to the treatment in section~\ref{subsec:low-energy}, we will first combine the excitations in a bispinor and then concentrate on the excitations close to the Fermi points. The results are independent of the  order of operations, but in this case it is simpler to proceed this way. Starting with the $\uparrow$ spin projection, the equations of motion are given by:
\begin{equation}
i\partial_t\left(\begin{array}{c}a_{\bm{p}+\bm{p}_\superfluid ,\uparrow}\\ a^\dagger_{-\bm{p}+\bm{p}_\superfluid ,\uparrow}\\\end{array}\right)=H_{\bm{p},\bm{p}_\superfluid ,\uparrow}\left(\begin{array}{c}a_{\bm{p}+\bm{p}_\superfluid ,\uparrow}\\ a^\dagger_{-\bm{p}+\bm{p}_\superfluid ,\uparrow}\\\end{array}\right),
\end{equation}
where
\begin{equation}
H_{\bm{p},\bm{p}_\superfluid ,\uparrow}:=M(\bm{p})\sigma_3+\bm{p}\cdot\bm{v}_\superfluid \,\sigma_0-c_\bot\bm{p}_m\sigma_1-c_\bot\bm{p}_n\sigma_2.
\end{equation}

Similar manipulations with the $\downarrow$ spin projection permit us to write the following evolution equation for bispinors:

\begin{equation}
i\partial_t\left(\begin{array}{c}a_{\bm{p}+\bm{p}_\superfluid ,\uparrow}\\ a^\dagger_{-\bm{p}+\bm{p}_\superfluid ,\uparrow}\\a_{\bm{p}+\bm{p}_\superfluid ,\downarrow}\\ a^\dagger_{-\bm{p}+\bm{p}_\superfluid ,\downarrow}\\\end{array}\right)=H_{\bm{p},\bm{p}_\superfluid }\left(\begin{array}{c}a_{\bm{p}+\bm{p}_\superfluid ,\uparrow}\\ a^\dagger_{-\bm{p}+\bm{p}_\superfluid ,\uparrow}\\a_{\bm{p}+\bm{p}_\superfluid ,\downarrow}\\ a^\dagger_{-\bm{p}+\bm{p}_\superfluid ,\downarrow}\\\end{array}\right),\label{eq:nonlinev}
\end{equation}
where the $4\times 4$ evolution operator is
\begin{equation}
H_{\bm{p},\bm{p}_\superfluid }:=M(\bm{p})Y^3+\bm{p}\cdot\bm{v}_\superfluid Y^0+c_\bot\bm{p}_m Y^1+c_\bot\bm{p}_n Y^2.
\end{equation}

The set of matrices $\{Y^1,Y^2,Y^3\}$ were defined in (\ref{eq:plamat}), and $Y^0:=I_{4}$. Now if we concentrate on excitations near the Fermi points ({\it i.e.} linearize around $\bm{p}=+p_\fermi\hat{\bm{l}}$; this linearization is sufficient as the equation is already encompassing all the degrees of freedom) one obtains a Dirac equation in momentum space:
\begin{equation}
\left(e^\mu_{\ I}\gamma^I \bar\mathfrak{p}_\mu+\gamma^0 p_\fermi\hat{\bm{l}}\cdot\bm{v}_\superfluid  \right)\psi_{\bm{p}}=0,\label{eq:nonlindir}
\label{eq:dirace}
\end{equation}
where $\psi_{\bm{p}}$ is the bispinor in (\ref{eq:nonlinev}), $\mathfrak p=\bm p- p_\fermi\hat{\bm{l}}$,  and $\bar\mathfrak{p}^\mu=(\omega,\mathfrak{p})$. The non-zero components of the tetrad $e^\mu_{\ I}$ are given by
\begin{eqnarray}
e^0_{\ 0}:=1~,~~~e^1_{\ 1}:=c_\bot ~,~~~e^2_{\ 2}:=c_\bot~,~~~
e^3_{\ 3}:=c_\|~,~~~e^i_{\ 0}:=v_\superfluid ^i~.
\end{eqnarray}
The corresponding metric components are
\begin{equation}
g^{\mu\nu}=\eta^{IJ}e^\mu_{\ I}e^\nu_{\ J}~,~~~\to~~~
g^{\mu\nu}=
\left(
\begin{array}{c|c}
-1 & -v_\superfluid ^i \\
\hline
-v_\superfluid ^i & D^{ij}-v_\superfluid ^i v_\superfluid ^j 
\end{array}
\right),
\label{eq:acousticmetric}
\end{equation}
with
\begin{equation}
D^{ij}=
\left(
\begin{array}{ccc}
c_\bot^2 & 0& 0\\
0& c_\bot^2 &0 \\
0&0 & c_\|^2
\end{array}
\right).
\end{equation}
This is an acoustic metric~\cite{Barcelo2005} which, given that the we are assuming a uniform background velocity $\bm{v}_\superfluid $, corresponds to a flat Minkowski spacetime. The equation (\ref{eq:dirace}) is completely equivalent to (\ref{eq:dirace0}) in the homogeneous case. 
The only difference is a constant shift in the energy of quasiparticles.
 
In order to discuss inhomogeneities it is better to work in position space. Eq. (\ref{eq:nonlindir}) would then be written as:
\begin{equation}
e^\mu_I\gamma^I(i\partial_\mu-B_\mu)\psi=0,\label{eq:nonlinevpos}
\end{equation}
where $\partial_\mu:=(\partial_t,\bm{\nabla})$ is the derivative operator including time, and $B_\mu:=(p_\fermi\hat{\bm{l}}\cdot\bm{v}_\superfluid ,p_\fermi\hat{\bm{l}})$ is a constant background. The content of this equation is then completely equivalent to one with $B_\mu =0$, since a constant background value can be absorbed into unobservable offsets of energy and momentum.
However, here we want to consider fluctuations of the background $\delta\hat{\bm{l}}$ and $\delta\bm{v}_\superfluid $. These fluctuations act as an effective vector field which affects the evolution of quasiparticles:
\begin{equation}
 e^\mu_{\ I}\gamma^I(i\partial_\mu -\nu  \bar{A}_\mu)\psi=0.\label{eq:diraceq1}
\end{equation}
Here $\nu$ is a constant which controls the dimensions of the field $\bar{A}_\mu$ (recall that in standard electrodynamics the vector potential has dimensions of momentum per unit charge).
The kind of coupling of the fermionic quasiparticles to the vector field $\bar{A}^\mu$ suggests that we identify it as an effective electromagnetic gauge field, as in other inhomogeneous situations in condensed matter physics (see {\it e.g.} \cite{Dalibard2011}). However, we should put the metric in its standard Minkowskian form before carrying out this identification. To do that, we  transform to comoving  coordinates so that the $v_\superfluid ^i \partial_i$ term in Dirac's equation (\ref{eq:diraceq1}) vanishes. Then the vector field is identified as 
\begin{equation}
\bar{\A}:= \frac{1}{\nu} \delta(p_\fermi\hat{\bm{l}}),\qquad \bar{A}_0=\frac{1}{\nu}p_\fermi\hat{\bm{l}}\cdot\delta\bm{v}_\superfluid .\label{eq:Adef}
\end{equation}
The object $\bar{\A}$ is a genuine vector, with three degrees of freedom: two originate from the variations $\delta \hat{\bm{l}}$ of the order parameter, and the other one from density fluctuations $\delta p_\fermi$. On the other hand, $\bar{A}_0$ contains just one degree of freedom independent of these.

To end this subsection, let us also point out that the inhomogeneities in the order parameter make the acoustic metric (\ref{eq:acousticmetric}) to be non-flat. Thus when considering higher-than-first-order effects, the same degrees of freedom making up this effective electromagnetic potential will be responsible for some partial curved-spacetime effects.

\subsection{Gauge symmetry and dynamics \label{sec:gauge}}

Our discussion so far has shown that the low-energy description of the system contains features that are not included in the original theory, such as the notion of electric charge and chirality. In this section we shall deal with another emergent property: gauge symmetry. When gauge fields are emergent entities, the discussion naturally splits in two aspects: on the one hand, the \emph{kinematical} invariance of the theory under gauge transformations and, on the other hand, the \emph{dynamical} preservation of this symmetry \cite{Jannes2009,Carlip2012}. The study of analogue gravity setups, where the relevant gauge group is composed by diffeomorphisms, has shown that condensed-matter analogies usually fail to achieve the second point \cite{Barcelo2005,Sindoni:2011ej}. This section is devoted to an analysis of these issues in the context of the model developed in this article, where the gauge group is simpler.

By kinematical gauge invariance we refer to a property of the way in which the low-energy quasiparticle excitations, the Nambu-Gor'kov spinors in (\ref{eq:ngup}) and (\ref{eq:ngdown}), react under the presence of different given fields $\bar{A}_\mu$, independent of their origin. As we have seen, the fields $\bar{A}_\mu$ are associated with spatial and temporal variations of the orbital part of the order parameter, which is represented by a trihedral $\{\hat{\bm{m}},\hat{\bm{n}},\hat{\bm{l}}\}$. Kinematic gauge invariance occurs when there are equivalent classes of $\bar{A}_\mu$ leading to essentially the same effect upon the quasiparticles. 

Consider as an example the following static texture:
\begin{equation}
\delta\hat{\bm{m}}=\delta\hat{\bm{m}}(\bm{x}\cdot\hat{\bm{m}}_0),\qquad \delta\hat{\bm{n}}=\delta\hat{\bm{n}}(\bm{x}\cdot\hat{\bm{n}}_0),\qquad \delta\bm{p}_\superfluid =0.\label{eq:text0}
\end{equation}
In this case, one can find a local phase transformation of the fermionic fields that transforms the evolution equation (\ref{eq:diraceq1}) into a free Dirac equation for the new spinor field. That is, for internal observers, the configuration (\ref{eq:text0}) would be equivalent to the absence of textures if they identify the physical objects with equivalence classes defined by these gauge transformations. A spinor field wave packet is not deflected in any way by the previous texture and one could take that as a defining feature of the equivalent class of configurations. As we could have anticipated, two textures differing in the gradient of a scalar, $\bar{A}'_\mu-\bar{A}_\mu=\partial_\mu \varphi$, lead to the same type of effects in the spinor field; the function $\varphi$ can always be absorbed locally into the spinor's local phase:
\begin{equation}
\psi\ \longrightarrow\ \exp[i\varphi(t,\bm{x})]\psi.
\end{equation}

Recall that, in the same way, in Maxwell's model (section \ref{sec:maxwell} and \cite{Maxwell1861a,Maxwell1861b}) electromagnetic potentials have also a reality but some of their properties are not relevant at low energies. 
At this point it is important to remark again that this picture is only partial: the description of the system is simple because we are looking only at low-energy phenomena. In particular, this gauge invariance will be violated at some point when the low-energy description breaks down, for instance, when the effective Lorentz invariance disappears. At some point even the condensation and so the very existence of the field $\bar{A}_\mu$ would disappear. Moreover, we are not considering the excitation of other collective modes, assuming that they are frozen. To derive the low-energy Dirac equation (\ref{eq:diraceq1}) with vector potential (\ref{eq:Adef}) we have assumed that these other collective modes are not excited (for example, the clapping modes which can be associated with gravitons; see \cite{Volovik1998} for a general discussion of the different collective modes and their significance, and \cite{Jannes:2011em} for the surprising relation between these clapping modes and the effective cosmological constant in $^3$He-A).

Let us now discuss the issue of dynamical gauge invariance. In principle it could be the case that the kinematical gauge invariance was not preserved by the dynamics. By looking at the interaction of two spinor wave packets, through a mediator field $\bar{A}_\mu$, one could detect differences beyond the introduction of a local phase. This amounts to the possibility of distinguishing between different members of the kinematical equivalence class. The emergence of a dynamical gauge invariance will definitely signal the irrelevance of certain degrees of freedom of $\bar{A}_\mu$ in the self-consistent low-energy physics of the system.

The obtention of a dynamical gauge invariance has turned out to be an issue much more subtle than expected. Following the literature, one is first naturally led to believe that the key of the question resides in an induction mechanism {\it \`a la} Sakharov \cite{Sakharov1968} (adapted to electromagnetism by Zel'dovich \cite{Zel'dovich1967}). However, we found this path to be paved with problems (for a summary of those problems, see \ref{sec:induction}). Rethinking the problem, we realised that the key may well reside in the very emergence of Lorentz invariance. Here we are going to describe the logic of the emergence of dynamical gauge invariance along this line of thought.

So far we have not inquired about the origin of the inhomogeneities in the order parameter, or in other words, of the $\bar{A}_\mu$ fields. For instance, a specific texture could be forced upon the system by using external forces. However, here we want to have a closed self-consistent system. Then, at low energy the inhomogeneities of the order parameter can only be induced by the presence of fermionic quasiparticles. For that it is useful to think of coherent states of fermionic quasiparticles, {\it i.e.} macroscopic spin waves. The source of $\bar{A}_\mu$ exhibits Lorentz invariance below the energy scale (\ref{eq:lorentz}). We have used the notation $\bar{A}_\mu$ to denote the objects that appear directly in Dirac's equation when written in comoving Lab coordinates. From the Lab perspective we know that $\bar{A}_i$ is a vector and $\bar{A}_0$ a scalar under spatial changes of coordinates. However, the solutions of this Dirac equation under the presence of $\bar{A}_\mu$ will be connected with those worked out in a Lorentz transformed coordinate frame but now under the presence of a Lorentz transformed object $A_\nu = \Lambda^\mu_\nu \bar{A}_\mu$. From the kinematic perspective, internal low-energy observers of quasiparticles are not able to pick out any special or privileged Minkowskian inertial frame. Thus they will construct their world view using an, in principle, generic object $A_\mu$ with the transformation properties of a Lorentz covariant four-vector.

However, one question is how the object $A_\mu$ transforms under a Lorentz change of coordinates, another question is whether the entire system is invariant under active Lorentz transformations or, in other words, whether the dynamics of $A_\mu$ is Lorentz invariant. As we said before, we are allowed to consider only those $A_\mu$ produced by the presence of quasiparticles. It is then a reasonable hypothesis that two Lorentz related sources lead to two Lorentz related $A_\mu$'s, {\it i.e.} that the Lorentz covariant structure of the spinor waves is passed over to the texture field. We do not have a proof of this reasonable hypothesis
but for the moment let us just assume that it is indeed valid in our system. Under this hypothesis, internal observers will write down a generic Lagrangian for the system of the form 
\begin{equation}
\mathscr{L}(\psi,A_\mu):=-\frac{1}{4\mu_0}{F}^{\mu\nu}F_{\mu\nu}+\frac{\xi}{2\mu_0}(\partial_\mu A^\mu)^2+\frac{m^2}{2\mu_0} A_\mu A^\mu+\mathscr{L}_{_{\rm{D}}}(\psi,A_\mu).\label{eq:effintlag}
\end{equation}
Let us first examine the possible value of the mass constant in the effective Lagrangian density (\ref{eq:effintlag}). A simple argument shows that $m$ cannot be different from zero. A non-zero value for the mass parameter would mean that to create a texture there would always have to be an energetic gap, no matter how smooth the texture may be ({\it i.e.}, no matter how large the associated wavelengths). But precisely these smooth variations of the order parameter explore the degeneracy manifold of the planar order parameter (\ref{eq:plaord}). Thus, in the limit of very long wavelengths, the construction of a texture should cost no energy. This is clear when we think of orbital textures, but it might appear less clear for perturbations of the superfluid velocity field. The problem is that from the point of view of the order parameter, a constant velocity already imposes a specific length scale for the variations of the phase. The perturbations of the velocity are encoded in second derivatives of the phase. If perturbations of the velocity field did have a gap, then, among other things, our Lorentz-invariant hypothesis would be broken as the $A_\mu$ would have an anisotropic mass (indeed, this is what happens in the A-phase of $^3$He when spin-orbit interactions are taken into account \cite{Volovik2003}). However, it is well known that assuming a fixed constant velocity background, the extra energy associated to the introduction of acoustic waves is such that its dispersion relation is gapless. So, indeed there is no mass term for $A_\mu$ in this theory.

The equations of motion for $m=0$ are
\begin{equation}
\square {A}_\mu-(1+\xi)\partial_\mu \partial_\nu{A}^\nu=j_\mu.
\end{equation}
The source of this equation of motion is the identically conserved fermionic current $j_\mu$. If one takes the divergence in the last equation, one finds for $\xi\neq0$ the following equation ($\xi=0$ leads to a trivial identity):
\begin{equation}
\square(\partial^\mu {A}_\mu)=0.
\end{equation}
In this way, the divergence $\partial_\mu A^\mu$ effectively behaves as a free scalar field, not coupled to the rest of fields (note that when gravity is included this is no longer true; in fact, the existence of such a scalar degree of freedom could have non-trivial cosmological implications \cite{Maroto2011}). The absence of sources for this construction makes it natural to impose a zero value of this field or, in other words, the Lorenz gauge condition
\begin{equation}
\partial^\mu {A}_\mu=0.
\end{equation}
Instead of working with this specific gauge fixing condition, one could choose to work  from the start with a theory without the $(\partial _\mu A^\mu)^2$ term in the Lagrangian. Then the theory will exhibit standard gauge invariance and one could proceed with any gauge fixing one likes. Both ways of proceeding will lead to the same physical results. Let us here make an interesting observation. To construct a fundamental theory of massless relativistic spin-1 particles, it is compulsory to introduce gauge invariance (see {\it e.g.}~\cite{Jenkins2006}). The $A_\mu$ field cannot be observable in a theory with a fundamental Lorentz invariance (unbroken at all energies). However, when Lorentz invariance is effective and appears only at low energies, the underlying theory can associate a physical reality to the $A_\mu$, but the low-energy observers are oblivious to some of its properties.\footnote[2]{It is interesting to notice that, in the case in which the gauge potential is identified with the four-vector describing the flow of matter, both the Lorenz and Coulomb gauge-fixing conditions can be physically interpreted in terms of properties of the underlying fluid ~\cite{Rousseaux2002,Rousseaux2005}. However, in this work we propose a different identification of the relevant vector gauge field at low energies.}

Regarding the value of the remaining constant, a dimensional analysis shows that the vacuum permeability would be given by the following expression:
\begin{equation}
\mu_0=\frac{4\pi m^*\hbar}{\nu^2p_\fermi}\alpha,
\end{equation}
where $\alpha$ is a dimensionless constant. The notation is not accidental: it corresponds to the effective fine-structure constant of the theory. Moreover, the only dimensionless quantity one can construct from the constants in the problem is the quotient $c_\bot/c_\|$. This means that the effective fine-structure constant must be calculable as a function of this quantity, {\it i.e.}
\begin{equation}
\alpha=\alpha\left(\frac{c_\bot}{c_\|}\right).
\end{equation}
This is all we can say with certainty about this quantity. Just like in effective field theories, the concrete form of the fine-structure constant must be determined by comparing a process ({\it e.g.} scattering of two quasiparticles) in both the low-energy theory and the condensed-matter theory with Hamiltonian (\ref{eq:hampairing}), in which all the constants are explicit. However, the occurrence of the condensation could hinder this comparison, as it implies a non-trivial resummation of the perturbative contributions at different orders.
This is beyond the scope of this paper, although it is certainly interesting. Notice that it is natural to expect a behaviour which guarantees that this scattering amplitude tends to zero when there is no interaction between fermions [$g\rightarrow 0$ limit in (\ref{eq:hampairing})].   

What about the very presence of Lorentz invariance in the $A_\mu$ sector?
We do not have a definitive argument that this should be the case, only some hints.
Imagine for example that the $A_\mu$ field could propagate faster than the effective speed of light as defined by the fermionic Lorentz symmetry. Then, the condensate would probably be unstable as it would be energetically favourable for the particles $\bm{p},-\bm{p}$ in the pair to become unpaired quasiparticles (producing some sort of Cherenkov radiation). Another argument is that the inhomogeneous perturbations of the condensed state might be seen as a coherent field of particle pairs moving on top of a homogeneous background condensate.
Within that interpretation it would be natural to expect that the velocity of these pairs would follow the same dispersion relation as their free particle cousins. 

In summary, if Lorentz invariance appears below the energy scale (\ref{eq:lorentz}), as it is tied up to the existence of Fermi points, the resulting dynamical theory would be gauge invariant and so indistinguishable from standard electromagnetism. From the perspective of electrodynamics as a fundamental theory, the imposition of Lorentz invariance and the fact that the interaction is mediated by a massless vector field which couples to the fermion current density are necessary and sufficient to obtain a gauge-invariant theory. This extends to the emergent scenario, thus fixing these as the relevant conditions one has to set up to completely reproduce electrodynamics at low energies. 
One can wonder whether this result, {\it i.e.} the secondary character of the principle of gauge invariance, is particular to electrodynamics or not.
In this respect, a detailed study of the non-abelian case and/or higher-spin fields (especially the spin-2 case) will be presented elsewhere.

\section{Summary and conclusions}

In this work we make a case for understanding electrodynamics from an emergent point of view. We do not commit with the specifics of the models presented. Quite on the contrary, we want to turn the emphasis to the generic characteristics of these models, those that would probably be shared by any emergent model of electrodynamics. We have also tried to convey the power contained in emergent constructions: very simple elementary components and interactions can lead to an enormously rich phenomenology as described by the effective theory.  

Although the constituents of both models are of different nature, it is not difficult to draw parallels. Maxwell's proposal contained two kinds of elements: vortical cells, whose most salient property is their ability to acquire rotation, and ball bearings, from which one constructs the analogue of charged matter. These two elements are also present in the model inspired by $^3$He. The role of movable ball bearings is now played by fermionic quasiparticles, low-energy excitations of a fundamental system of fermions subject to particular kinds of interactions. These low-energy excitations, or quasiparticles, evolve following Dirac's equation. On the other hand, when the fundamental fermions are paired up and condensed they act as vortical cells, which possess intrinsic rotational properties due to the finite value of angular momentum characteristic of the $p$-wave condensation. The electromagnetic fields analysed here arise as the coarse-grained view of these effective bosons, {\it i.e.} as perturbations of the condensed phase.

In both models the velocity of light is emergent. Since both theories have been formulated as Galilean theories, there is in principle no obstruction for the elementary component to travel at arbitrarily large velocities. The speed of light appears as a ``sound'' speed, the velocity of wave-like excitations in the system. In the case of the superfluid model this velocity and its independence of the wavelength is strongly tied up to the occurrence of Fermi points where the dispersion relation becomes relativistic. All the physics could be described by a privileged external observer by using Newtonian notions. However, internal low-energy observers would tend to develop ways to understand their low-energy world that do not assume external structures. This epistemological choice is certainly valid, but it seems that the price to pay would be the necessary assumption of some features as fundamental principles, and hence a loss of explanatory power.

Another interesting parallelism is that, in both cases, the electromagnetic potential has a physical reality in terms of specific properties of the system under consideration. It is only at low energies that some of these degrees of freedom become effectively invisible and the internal gauge symmetry appears.

Beyond these parallelisms, the superfluid model goes further than Maxwell's model.
\begin{enumerate}
\item
While in Maxwell's construction the two substances making up the system, charged matter and electromagnetic fields, are independently postulated, in the superfluid framework they arise from the same single set of underlying elements.

\item 
The superfluid model can take account of the spinorial and quantum-mechanical properties of matter. The notion of charge cannot emerge here from the (quantum mechanical) quasiparticle density that is always non-negative ($\psi^+ \psi \geq 0$). However, the fact that there exist two signs for the charge is a nice logical consequence of the appearance in pairs of the Fermi points.

\item
Moreover, it seems possible to include quantum features of the electromagnetic field in the $^3$He-like model. Individual photons would correspond to tiny fluctuations of the condensed phase -- so tiny that they involve only one of these effective bosons composed by a pair of fermions. The picture suggested by this model is that photons should not be viewed as fundamental particles, but as composite structures emerging from the same fundamental ingredients as the fermionic quasiparticles (there are other examples in the literature in which photons and electrons arise from the same underlying system, although in those constructions even Fermi-Dirac statistics is emergent~\cite{Wen2001}). At this point, this is only a (natural) conjecture but in future work we plan to analyse to what extent this can be rigorously formulated.

\end{enumerate}

Emergent views of the kind analysed in this paper always imply that the low-energy properties, for instance Lorentz invariance, will eventually break up at some high-energy scale. Thus, it is important to stress here that deviations from Lorentz invariance need not occur at the Planck scale (and indeed Lorentz violations at the Planck scale are almost excluded by experimental observations; see {\it e.g.} \cite{Maccione2009,Liberati2013}). Quite on the contrary, there are strong arguments suggesting that, if general relativity is an emergent theory, then Lorentz symmetry has to be very accurately respected at the Planck scale~\cite{Volovik2006,Barcelo2010}, and the characteristic energy scale of Lorentz symmetry breaking should have to be several orders of magnitude higher than the Planck scale. 
We intend to approach the gravitational emergent problem in a future work. 

Although the models presented here are Newtonian at high energies, we are far from suggesting that high-energy physics should be Newtonian. What these examples show is that high-energy physics will most probably incorporate ingredients rather distinct from those of its low-energy incarnation. The emergent perspective is capable of providing tantalizing explanations of principles of physics without relaying on the specifics of the high-energy theory. We thus think that, in our search of a deeper understanding of nature, an emergent point of view is a useful and probably even necessary complement to an analysis based on fundamental principles.

\ack
Financial support was provided by the Spanish MICINN through the projects FIS2011-30145-C03-01 and FIS2011-30145-C03-02 (with FEDER contribution), and by the Junta de Andaluc\'{\i}a through the project FQM219. R. C-R. acknowledge support from CSIC through the JAE-predoc program, cofunded by FSE. 

\appendix

\section[\hspace*{12ex}Ginzburg-Landau minimization]{Ginzburg-Landau minimization} 
\label{sec:GL}  

The validity of Ginzburg-Landau theory is restricted to temperatures near the critical  transition temperature $T_\critical$. However, it is much  easier to handle the calculations within this restricted setup, and then generalize them to the whole range of temperature by using the microscopic theory. In either case the structure of the order parameter (\ref{eq:orderpar}) is obtained by a minimization principle. In the microscopic theory the quantity to be minimised is the expectation value of the Hamiltonian (\ref{eq:hampairing}) in the corresponding Fock vacuum state. In Ginzburg-Landau theory (for constant temperature $T$ and volume $V$)   the Helmholtz free energy functional of the order parameter, constructed as follows, is the quantity to be minimised. The order parameter (\ref{eq:orderpar}) is zero above a certain critical temperature $T_\critical $ but takes a finite value for $T<T_\critical $.  If we suppose that, near $T_\critical $, the free energy is analytic in the order parameter and obeys the symmetries of the microscopic Hamiltonian, then one can write a Taylor expansion near the critical temperature. To have a non-trivial minimization problem of this free energy one only needs to take into account the first two non-zero orders. In our case, as the free energy must be invariant under rotations in both coordinate and spin spaces, these terms will be second-order and fourth-order. Given the order parameter (\ref{eq:orderparameter}), there is one possible second-order term and five fourth-order terms:
\begin{eqnarray}
I_0:=\sum_{ai}d_{ai}d^*_{ai},
 \\
I_1:=\sum_{ai}\sum_{bj}d_{ai}d_{ai}d^*_{bj}d^*_{bj},
 \\
I_2:=\sum_{ai}\sum_{bj}d_{ai}d_{bj}d^*_{ai}d_{bj}^*,
 \\
I_3:=\sum_{ai}\sum_{bj}d_{ai}d_{aj}d^*_{bi}d^*_{bj},
 \\
I_4:=\sum_{ai}\sum_{bj}d_{ai}d_{bj}d^*_{aj}d^*_{bi},
\\
I_5:=\sum_{ai}\sum_{bj}d_{ai}d_{bi}d^*_{aj}d^*_{bj}.
\end{eqnarray}
Then the general form of the free energy is
\begin{equation}
F=F_{\rm n}+\alpha_0(T-T_\critical )I_0+\frac{1}{2}\beta(T_\critical )\sum_{s=1}^5\beta_s I_s.
\end{equation}
Terms proportional to gradients of the order parameter are neglected because the variations of the order parameter are considered smooth enough. $F_{\rm n}$ is the free energy of the normal phase, which is independent of the order parameter so it is an irrelevant constant to our purposes. The form of the coefficients we are considering is enforced by the behaviour of the order parameter near the critical temperature (see Sec. 5.7 in \cite{Leggett2006} for a detailed discussion).

As it stands, this minimization problem is not analytically solvable. For this reason, the so-called unitarity condition,
\begin{equation}
\sum_{b,c}\epsilon_{abc}{d}^*_{bi} d_{cj}=0,
\end{equation}
is imposed. Although there is no theoretical argument to impose this condition (apart from simplicity of certain expressions), there is some experimental justification since it seems that the states of $^3$He which are realised in nature are all unitary in this sense. Consideration of non-unitary states could of course be interesting for other purposes. In our case, although the state we are most interested in (the planar state) does not seem to be realised in nature, it is nevertheless also unitary.

\section[\hspace*{12ex}Some comments regarding the ABM order parameter]{Some comments regarding the ABM order parameter} 
\label{sec:aboutABM}  

In this appendix we will make a brief analysis of the quasiparticle evolution equations for the ABM state with order parameter (\ref{eq:abmord}) and
\begin{equation}
\hat{\bm{x}}=\hat{\bm{m}},\qquad
\hat{\bm{y}}=\hat{\bm{n}}',\qquad
\hat{\bm{z}}=\hat{\bm{s}},\label{eq:axes}
\end{equation}
along the lines of the analysis performed for the planar state in the main text, to show the differences between both states. The reason for the choice of axes (\ref{eq:axes}) is that the alignment of $\hat{\bm{s}}$ and $\hat{\bm{m}}\times\hat{\bm{n}}$ is favoured by the action of nuclear dipole interactions \cite{Woelfle1976,Vollhardt1997}. The arguments and conclusions in this section do  not depend on this choice (see footnote below).

Because of this choice of axes, it is better to write the evolution equations of quasiparticle operators in the spin basis in $\hat{\bm{x}}$ direction, defined as
\begin{equation}
a_{\bm{p}\rightarrow}:=\frac{a_{\bm{p}\uparrow}+a_{\bm{p}\downarrow}}{\sqrt{2}},\qquad a_{\bm{p}\leftarrow}:=\frac{a_{\bm{p}\uparrow}-a_{\bm{p}\downarrow}}{\sqrt{2}}.
\end{equation}
When these equations are linearised around the Fermi point $+p_\fermi\hat{\bm{l}}$ and represented in position space, one obtains the analogue of Eq. (\ref{eq:nonlinevpos}) but split for both spin projections. One can check that one obtains similar equations directly in the basis $\uparrow$, $\downarrow$ when $\hat{\bm{s}}=\hat{\bm{x}}$. If, instead of that, one takes $\hat{\bm{s}}=\hat{\bm{y}}$, the equations are almost the same, with only a change of the sign of the two last terms in the last equation.  The evolution operators are given by the following expressions:
\begin{eqnarray}
\mathscr{H}_\rightarrow:=c_\|\hat{\bm{l}}\cdot(-i\bm{\nabla}-p_\fermi\hat{\bm{l}})
+c_\bot(\sigma_1\hat{\bm{m}} - \sigma_2\hat{\bm{n}})\cdot(-i\bm{\nabla}),
\nonumber\\
\mathscr{H}_\leftarrow:=c_\|\hat{\bm{l}}\cdot(-i\bm{\nabla}-p_\fermi\hat{\bm{l}})
-c_\bot(\sigma_1\hat{\bm{m}} + \sigma_2\hat{\bm{n}})\cdot(-i\bm{\nabla}).
\end{eqnarray}
One can realize that these equations both have the same chirality, by multiplying the $\pm1$ factors in front of the Pauli matrices. For this reason it is better to apply a linear transformation to one of the equations, say the second, to change its chirality. Such a transformation is given by
\begin{equation}
\psi_\leftarrow\ \longrightarrow\ i\sigma_2\psi_\leftarrow^*.
\end{equation}
The transformed Hamiltonian is
\begin{equation}
\mathscr{H}_{\leftarrow}':=-\sigma_2\mathscr{H}_{\leftarrow}^*\sigma_2=-c_\|\hat{\bm{l}}\cdot(-i\bm{\nabla}+p_\fermi\hat{\bm{l}})
+c_\bot(\sigma_1\hat{\bm{m}} - \sigma_2\hat{\bm{n}})\cdot(-i\bm{\nabla}).
\end{equation}
In the same way as we did with the planar phase, let us define the matrices
\begin{equation}
Z^1=\left(\begin{array}{cc}\sigma_1&0\\0&\sigma_1\\\end{array}\right),\quad Z^2=\left(\begin{array}{cc}-\sigma_2&0\\0&-\sigma_2\\\end{array}\right),\quad Z^3=\left(\begin{array}{cc}\sigma_3&0\\0&-\sigma_3\\\end{array}\right).
\end{equation}
The problem of finding a representation of the gamma matrices $\{P,PZ^a\}_{a=1,2,3}$ is similar to the one studied for the case of the planar state. Here, a solution is given by
\begin{equation}
P:=\left(\begin{array}{cc}0&\sigma_3\\\sigma_3&0\\\end{array}\right).
\end{equation}
However, now the two chiralities have a different coupling to the vector potential when the same perturbative analysis of Sec. \ref{sec:text} is applied to the evolution equations of this section. Such a coupling implies that we cannot write the low-energy evolution equations as a Dirac field coupled to a vector potential. In fact, the coupling is now axial, with a term in the equation of motion proportional to
\begin{equation}
\gamma^5 \gamma^\mu\bar{V}_\mu\psi.
\end{equation}
Here we are using the symbol $\bar{V}_\mu$ instead of $\bar{A}_\mu$ to denote the inhomogeneities of the orbital part of the order parameter. The reason is that, even if these objects are written in the same way in terms of the inhomogeneities, they should have different transformation properties under the usual symmetry transformations such as parity. This can be understood by looking at the structure of the Cooper pairs in both states, ABM and planar. In the first case the vector $\hat{\bm{l}}$ shows the direction of the angular momentum of the Cooper pairs with positive as well as negative projection of spin, and so, $\hat{\bm{l}}$ is an axial vector in this state. On the other hand, in the planar state the two spin populations form Cooper pairs with opposite angular momentum, implying that the planar state is not axial (for a more detailed discussion see Sec. 7.4 of \cite{Volovik2003}). This picture is consistent with the kind of coupling to a vector field which appears in each state.

This does not contradict the claim that the low-energy quasiparticle excitations of the Fermi liquid are determined by topology in momentum space. Both ABM and planar states are characterised by two Fermi points. In a homogeneous system, the low-energy fermionic excitations are the same in both states, and can be represented by a (free) Dirac field. However, the structure of the order parameter is different in both states, and so is the coupling of the fermionic excitations to the inhomogeneities of this order parameter. In other words, the only difference between these two states is their chirality. This is the reason why the planar state serves better as an analogue of  the vacuum state of electrodynamics.

ABM and planar states are limiting cases of the family of axiplanar states \cite{Volovik2003}. In these states the two spin populations are decoupled as far as the order parameter is concerned. For these two limiting states, one is considering perturbations with $\delta\hat{\bm{l}}_\uparrow=\delta\hat{\bm{l}}_\downarrow$ and $\delta\hat{\bm{l}}_\uparrow=-\delta\hat{\bm{l}}_\downarrow$, respectively. General axiplanar states can be analysed with the same techniques to show that, in general, one has couplings to a polar as well as an axial vector, both constructed by different linear combinations of the independent variations $\delta\hat{\bm{l}}_\uparrow$ and $\delta\hat{\bm{l}}_\downarrow$, {\it i.e.} $(\delta\hat{\bm{l}}_\uparrow\pm\delta\hat{\bm{l}}_\downarrow)/\sqrt{2}$. 


\section[\hspace*{12ex}Inhomogeneous Ginzburg-Landau free energy and Zel'dovich picture of emergence]{Inhomogeneous Ginzburg-Landau free energy and Zel'dovich picture of emergence} \label{sec:induction}

In our discussion, we have not followed the usual way in which the issue of the dynamics of the order parameter is approached (see {\it e.g.} \cite{Volovik2003}). Let us shortly describe it, as well as its shortcomings.

From a condensed-matter perspective, it is natural to expect that the dynamics of the order parameter can be determined by a generalised Ginzburg-Landau approach, analogue to the one sketched in Sec. \ref{sec:order}, but this time retaining terms containing derivatives of the order parameter. In a similar way, the temperature-dependent coefficients accompanying each term of the free action can be calculated from the microscopic theory. The result of this calculation would be an expression quadratic in the derivatives of the textures. If, for the moment, we restrict the discussion to orbital textures of the unit vector, $\delta\hat{\bm{l}}$, the corresponding inhomogeneous part of the free action was worked out in \cite{Cross1975}. At finite temperature $T$ it is given by the following expression:
\begin{equation}
\frac{p_\fermi c_\|}{12\pi^2\hbar}\left[\log\left(\frac{\Delta_0}{k_\boltzmann T}\right)[\hat{\bm{l}}\times(\bm{\nabla}\times\hat{\bm{l}})]^2+[\hat{\bm{l}}\cdot(\bm{\nabla}\times\hat{\bm{l}})]^2+(\bm{\nabla}\cdot\hat{\bm{l}})^2\right].\label{eq:fetext1}
\end{equation}
In this expression we are keeping the dominant terms in the zero-temperature limit $T\rightarrow0$, as well as the first order in an expansion in the parameter $c_\bot/c_\|$. This expansion  is usually carried out in the literature because of the smallness of this parameter in the experimental case of $^3$He. The reason for the infrared divergence in the first term can be traced back to the existence of Fermi points in the fermionic spectrum. In laboratory realizations the infrared divergence is always regulated by the temperature of the system. However, the other terms have coefficients which are constant in the limit $T\rightarrow0$. Therefore one can always, in principle, lower the temperature sufficiently to make the first term in (\ref{eq:fetext1}) the dominant one. For completeness, let us mention an assumption one can read in the literature, concerning the existence of the following additional term in this expansion:
\begin{equation}
\frac{p_\fermi c_\|}{12\pi^2\hbar}\left(\frac{c_\bot}{c_\|}\right)^2\log\left(\frac{\Delta_0}{k_\boltzmann T}\right)[\hat{\bm{l}}\cdot(\bm{\nabla}\times\hat{\bm{l}})]^2.\label{eq:fetext2}
\end{equation}
It has been claimed \cite{Volovik2003} that this term has been usually neglected because it is of quadratic order in the parameter $c_\bot/c_\|$, but that it appears in an explicit evaluation at this order \cite{Dziarmaga2002}. From our point of view, however, there is no conclusive argument in this respect, as the definite relation between the evaluation of the Ginzburg-Landau energy and the technical procedure used in \cite{Dziarmaga2002} (which indeed seems to be closer to Zel'dovich's approach we discuss in this section) is not clear for us. Notice that the terms presented here, (\ref{eq:fetext1}) plus (\ref{eq:fetext2}), would correspond to the potential energy of a theory with the usual kinetic term for the restricted kind of textures considered, $(\partial_t\delta\hat{\bm{l}})^2$. In the following we will check whether or not all these terms can be obtained in a simpler way from the perspective of the emergent relativistic theory we have been describing in the main text. 

Within the emergent relativistic field theory framework, a possible way in which an internal observer can determine the dynamics of the gauge fields is by integrating out fluctuations of the relativistic fermionic fields. This is nothing but the suggestion of Sakharov concerning gravity \cite{Sakharov1968} (see \cite{Visser2002} for a modern review), adapted to electrodynamics by Zel'dovich \cite{Zel'dovich1967}. The integration over fermionic fluctuations, which technically amounts to an evaluation of a fermionic path integral in the presence of background fields, can be found in the literature carried out in different ways; see for example \cite{Andreev1987,Dziarmaga2002}.  A note of caution: these two approaches (Ginzburg-Landau on the one hand and Sakharov-Zel'dovich on the other) are, in principle, very distinct in nature. One is a finite-temperature analysis (implying that we have a thermal distribution of fermionic quasiparticles) while the second is a zero-temperature calculation. We will proceed in the comparison anyway, but keeping this in the back of our minds.

The only calculation we need is the evaluation of the one-loop polarization tensor, which characterizes the only divergent term in the fermionic path integral. This is a standard calculation which involves an ultraviolet regularization of the momentum integral with an upper cutoff $\Lambda_+$ as well as an infrared regularization by means of a similar quantity $\Lambda_-$. At this stage the significance of these quantities is merely formal, although they will gain a physical interpretation later. In the case in which Lorentz and gauge symmetries are preserved by the regularization method, the divergence is logarithmic in the limit $\Lambda_+/\Lambda_-\rightarrow\infty$ and corresponds to a term in the action 
\begin{equation}
-\frac{\nu^2}{48\pi^2\hbar}\log\left(\frac{\Lambda_+}{\Lambda_-}\right)^2\int\mbox{d}^4x\,F^{\mu\nu}F_{\mu\nu}.\label{eq:lag3a}
\end{equation}
In standard quantum field theory, this term would be absorbed by a suitable counterterm before taking the limit $\Lambda_+/\Lambda_-\rightarrow\infty$, leading to charge and photon field renormalization. 
However, in this construction we have no definite tree-level kinetic term for the vector potential. Moreover, the ultraviolet divergence here is clearly an artefact of the extrapolation of the low-energy theory to higher energies. Following Zel'dovich, we can interpret this (finite) term as the leading kinetic term for the electromagnetic field induced at one-loop level:
\begin{equation}
-\frac{1}{4\mu_0}\int\mbox{d}^4x\,F^{\mu\nu}F_{\mu\nu}.\label{eq:lag3}
\end{equation}
In principle, the logarithmic behaviour supports this interpretation since it permits this term to be dominant over any unknown, tree-level dynamical terms for the gauge field, at least for certain values of the quotient $\Lambda_+/\Lambda_-$. This is what would be usually understood as one-loop dominance \cite{Sakharov1968,Visser2002}. In this regime it is reasonable to think that the dynamics of the order parameter should be well described by (\ref{eq:lag3}). In the resulting effective theory, as expected, the actual value of $\nu$ is not relevant since it can be changed by a redefinition of the vector potential; the only meaningful quantity is the combination
\begin{equation}
\alpha:=\frac{\nu^2\mu_0}{4\pi\hbar}=\frac{3\pi}{\log(\Lambda_+/\Lambda_-)}.\label{eq:ccons}
\end{equation}
Notice that the value of this dimensionless coupling constant, the effective fine-structure constant, is universal: it only depends on the value of the ultraviolet and infrared cutoffs.

The next step would be the comparison of the term (\ref{eq:lag3a}) with the corresponding limit of the Ginzburg-Landau approach, which contains in principle all the information about the evolution of textures. We are going to do it for restricted textures in which the superfluid velocity plays no role. At low temperatures two terms in (\ref{eq:fetext1}) and (\ref{eq:fetext2}) are the dominant ones. At first order in the texture, for which
\begin{equation}
-\frac{\nu}{p_\fermi}\bm{A}=\delta\hat{\bm{l}}=\delta\hat{\bm{m}}\times\hat{\bm{n}}_0+\hat{\bm{m}}_0\times\delta\hat{\bm{n}},
\end{equation}
one can see that the sum of these terms is equivalent to the spatial part of the relativistic term which can be written as:
\begin{eqnarray}
\frac{1}{2}F^{\mu\nu}F_{\mu\nu}=& 
 c_\bot^2(\partial_0A_1)^2+c_\bot^2(\partial_0A_2)^2-c_\bot^4(\partial_1A_2-\partial_2A_1)^2
\nonumber\\&-c_\bot^2c_\|^2(\partial_3A_1)^2-c_\bot^2c_\|^2(\partial_3A_2)^2.
\end{eqnarray}
In this way, one can in principle accept that the picture of induction of dynamics captures the relevant dynamics of the system when the temperature is low enough. As long as Lorentz invariance is kept intact in the present scheme, one can argue in favour of the occurrence of the term (\ref{eq:fetext2}) in the inhomogeneous Ginzburg-Landau free energy. Then the matching of the low-energy relativistic theory with the Ginzburg-Landau approach tells us the value of the quotient of regulators:
\begin{equation}
\frac{\Lambda_+}{\Lambda_-}=\frac{\Delta_0}{k_\boltzmann T}.
\end{equation}
This fixes the value of the fine-structure constant (\ref{eq:ccons}) in terms of parameters of the system and provides an interpretation of the ultraviolet and infrared regulators. They would be given by:
\begin{equation}
\Lambda_+\simeq \Delta_0,\qquad \Lambda_-\simeq k_\boltzmann T.
\end{equation}
The value of the infrared regulator simply implies that the energy scale associated with the temperature physically removes the infrared divergence. On the other hand, the value of the ultraviolet regulator is telling us that we are performing the integration over fermions up to energies given by $\Delta_0$. The trouble with this observation is that this energy is much greater than the scale of violation of Lorentz symmetry $E_\lorentz=m_*c_\bot^2$ [recall the discussion around Eq. (\ref{eq:lorentz})]:
\begin{equation}
\frac{\Delta_0}{m^*c_\bot^2}=\frac{c_\|}{c_\bot}\gg1.
\end{equation}
Thus we have no definite argument which supports that the dynamics should be given by the standard relativistic term (\ref{eq:lag3}), and the whole argument falls apart. The only way to remedy this would be to evaluate the inhomogeneous Ginzburg-Landau free energy in order to check the occurrence of (\ref{eq:fetext2}). It is interesting to notice that there exists a different scale of violation of Lorentz symmetry for quasiparticles travelling along the anisotropy axis (only $\frak{p}_l\neq0$ in Eq. \ref{eq:disp}) with higher characteristic energy, which could help to obtain this result. However, even if one succeeds on this, the picture would have additional negative features as we argue in what follows.

In the Ginzburg-Landau approach there are additional terms which do not respect the low-energy emergent symmetries, relativistic and gauge invariance. On the one hand, higher orders in the logarithmic term (109) are suppressed by inverse powers of the Fermi momentum, which is several
orders of magnitude greater than the maximum momentum which can be carried by the electromagnetic field (recall Eq. (\ref{eq:kmax})). Notice that gauge invariance is a property only of the linearised version of (\ref{eq:fetext1},\ref{eq:fetext2}). On the other hand, terms which are quadratic in the texture but non-relativistic (the relevant symmetry group is the Galileo group) are suppressed in a weaker way by the logarithm in (\ref{eq:fetext1},\ref{eq:fetext2}) or, equivalently, (\ref{eq:lag3a}) at low temperatures. These terms come from the gradient expansion of the Goldstone variables so they are ultimately linked to the breaking of symmetries in the condensed phase. The first unsatisfactory feature of this argument is that, since this suppression is logarithmic, one would have to consider practically zero temperatures to ensure that the logarithm is large enough. But there is even a stronger argument against this picture: the value of the fine-structure constant (\ref{eq:ccons}) shows that the suppression of these terms is proportional to $\alpha$. Thus in this effective theory the usual perturbative expansion in terms of the fine-structure constant makes no sense.

All these arguments make it difficult to support a comparison between the Ginzburg-Landau free energy and the low-energy action for the emergent gauge fields. The mechanism which permits to obtain a dynamical implementation of gauge invariance cannot be a logarithmic suppression of the terms which violate this symmetry in the action (it is also difficult to reconcile this logarithmic suppression with the accuracy of known symmetries \cite{Iliopoulos1980}). Our arguments are sufficiently general to apply to the original discussion of Zel'dovich \cite{Zel'dovich1967}, as it will be discussed elsewhere.

\newpage

\section*{References}

\bibliography{electromagnetism-emergence}

\end{document}